\documentclass[twocolumn, switch]{article} 
\usepackage{preprint}
\usepackage{algorithm}
\usepackage{algorithmic}
\usepackage{amsmath, amsthm, amssymb, amsfonts}

\usepackage[numbers,square]{natbib}
\usepackage[utf8]{inputenc}	
\usepackage[T1]{fontenc}	
\usepackage{xcolor}		
\usepackage[colorlinks = true,
            linkcolor = cyan,
            urlcolor  = blue,
            citecolor = cyan,
            anchorcolor = black]{hyperref}	
\usepackage{booktabs} 		
\usepackage{nicefrac}		
\usepackage{microtype}		
\usepackage{lineno}		
\usepackage{float}			

\usepackage{afterpage}
\usepackage{dblfloatfix}
\usepackage{caption}
\usepackage{multirow}
\usepackage{tabularx}
\usepackage{longtable}
\usepackage{array}
\usepackage[normalem]{ulem}
\usepackage{tcolorbox}
\tcbuselibrary{breakable, skins}
\newtcolorbox{promptbox}[1][]{
  colback=gray!5,
  colframe=gray!50,
  fonttitle=\bfseries\small,
  breakable,
  left=6pt, right=6pt, top=4pt, bottom=4pt,
  title={#1}
}
\newcounter{supptext}
\renewcommand{\thesupptext}{S\arabic{supptext}}
\newcommand{\supptextcaption}[1]{%
  \refstepcounter{supptext}%
  \noindent\textbf{Supplementary Text \thesupptext\,:} #1\par\vspace{6pt}%
}

\usepackage{titlesec}
\titlespacing\section{0pt}{12pt plus 3pt minus 3pt}{1pt plus 1pt minus 1pt}
\titlespacing\subsection{0pt}{10pt plus 3pt minus 3pt}{1pt plus 1pt minus 1pt}
\titlespacing\subsubsection{0pt}{8pt plus 3pt minus 3pt}{1pt plus 1pt minus 1pt}

\title{Statistical realism is not evidence that LLMs can estimate treatment effects in social science experiments
}

\usepackage{eso-pic}
\usepackage{tikz}
\usepackage{xcolor}
\PassOptionsToPackage{colorlinks=true,linkcolor=gray,urlcolor=gray}{hyperref}

\newcommand{\AddMyWatermarks}{%
  \begin{tikzpicture}[remember picture, overlay]
  \end{tikzpicture}%
}

\AddToShipoutPictureBG*{\AddMyWatermarks}

\usepackage{titling}
\usepackage{orcidlink}
\usepackage{footmisc}
\newcommand{\Author}[3]{
  \textbf{#1}\textsuperscript{#2}\ \orcidlink{#3} %
}

\author{
  \Author{Zonghan Li}{1}{0000-0003-0253-0139} \and
  \Author{Feng Ji}{1}{0000-0002-2051-5453}
}
\date{%
  \textsuperscript{1}Department of Applied Psychology and Human Development, University of Toronto, ON, Canada\\
  [4pt]
  \footnotesize \textbf{Corresponding author:} Feng Ji \texttt{<f.ji@utoronto.ca>}\\
}

\begin{document}

\twocolumn[ 
  \begin{@twocolumnfalse} 

\maketitle
\thispagestyle{empty}

\begin{abstract}

Large language models (LLMs) are increasingly used to simulate human responses and estimate treatment effect of interventions when real-world experiments are costly or infeasible. The treatment-effect estimates are often evaluated using statistical realism, the degree to which simulated responses reproduce properties of observed human responses, although whether realism predicts treatment-effect accuracy remains unknown. Here we test this proxy relationship by jointly measuring statistical realism and treatment-effect accuracy on the same simulated responses in a cross-national experiment with 59,508 participants from 62 countries using three LLMs. The correlation between statistical realism and treatment-effect accuracy is weak, and optimizing for statistical realism can even worsen treatment-effect accuracy when selecting models, prompts, and target populations. The pattern replicates in two additional cross-national experiments spanning 12 and 27 countries with 20,785 participants. The divergence between the two reflects distinct error structures and is larger for behavioral outcomes, where models appear to extrapolate behavioral effects from attitudinal patterns. Because this divergence may remain hidden in deployment, errors can propagate into simulation-informed decisions. We introduce a diagnostic framework for LLM-generated synthetic data and discuss how treatment-effect validation should proceed under varying availability of experimental benchmarks. Simulated responses and simulated treatment effects are distinct estimation targets, and evidence for one does not certify the other.

\end{abstract}
\vspace{0.05cm}
\vspace{0.5cm}

  \end{@twocolumnfalse} 
] 



\section*{Main}

\begin{figure}
  \centering
  \includegraphics[width=\linewidth]{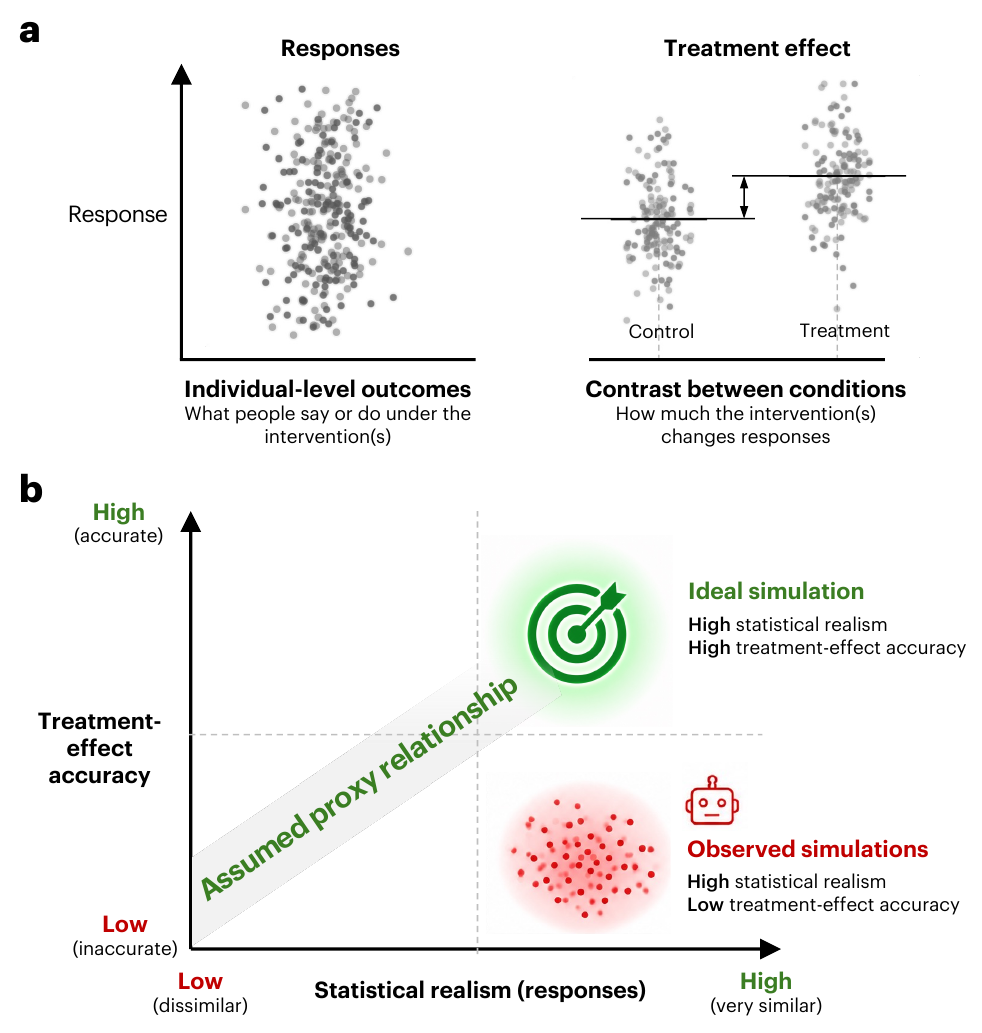}
  \caption{\textbf{Statistical realism and treatment-effect accuracy are distinct evaluation targets that can diverge.} \textbf{a.} The two targets are defined over different objects. A response is what a respondent says or does under the intervention(s) (left). A treatment effect is the change in responses produced by an intervention, obtained by contrasting conditions (right). Statistical realism evaluates the responses, and treatment-effect accuracy evaluates the contrast. \textbf{b.} Validation in practice assumes a proxy relationship between the two criteria (diagonal), in which higher realism indicates more accurate effect estimates. Observed simulations can instead fall in the lower-right region, where they pass realism-based validation while misestimating treatment effects.}
  \label{fig:fig1}
\end{figure}

Simulating how people will respond to interventions is a central task in the social and behavioral sciences \cite{RN1, RN2}. Large language models (LLMs) hold the potential to generate synthetic respondents in place of human participants \cite{RN3}, and are extending research to populations and settings that are costly, difficult or unethical to study directly \cite{RN4,RN5}. Emerging applications such as policy-message pretesting \cite{RN6,RN7,RN8}, generative digital twins \cite{RN9,RN10}, and agent societies \cite{RN11,RN12} are moving from the simulated responses themselves to simulation-derived estimates of treatment effects, the contrast between responses under an intervention and responses without it (Fig.~\ref{fig:fig1}a). As simulation-derived treatment effect estimates are increasingly discussed as evidence for research and policy design, pressing questions have emerged about whether and when they can be trusted.

The validation available in practice targets a different quantity from the one these applications seek to deliver. Direct validation of simulation-derived treatment effects requires observed experimental effects, which are often unavailable in the intended application settings. Validation has therefore rested largely on statistical realism \cite{RN13}, a response-level criterion that can be assessed without an experimental effect benchmark by testing whether model outputs reproduce properties of observed human responses \cite{RN14}, such as distributions \cite{RN14,RN15}, means \cite{RN16,RN17}, and group differences \cite{RN17,RN18}. Yet matching responses within conditions does not necessarily establish that a simulation recovers the treatment-effect contrast between them.

Existing comparisons with human experiments indicate that statistical realism and treatment-effect accuracy can diverge. Simulations can match aggregate statistics while compressing individual-level variation \cite{RN19}, recover the rank \cite{RN3} and direction \cite{RN3,RN20,RN21} of effects while inflating their magnitudes, and produce accurate effect estimates only in particular settings \cite{RN22,RN23}. A simulation may therefore pass statistical realism-based validations while still misestimating treatment effects (Fig.~\ref{fig:fig1}b), and the conclusions, policies and messages it informs may inherit errors that statistical realism-based checks do not detect. Existing studies, however, have not tested whether simulations with greater statistical realism also produce more accurate treatment-effect estimates, leaving uncertain whether response-level evidence supports effect-level claims.

Testing this proxy requires measuring statistical realism and treatment-effect accuracy in the same set of simulations against interventions with known effects. We therefore present realized interventions to the models as if their outcomes were unknown. In real-world experiments, treatment effect size varies with the composition of the population \cite{RN24}, intervention mechanism \cite{RN25}, and outcome type (e.g., an attitude or an actual behavior) \cite{RN26,RN27}, so a convincing test must span these sources of variation. Climate attitudes and behaviors provide such a testbed. Climate-related interventions draw on a wide range of mechanisms, the attitudes and behaviors involved differ markedly across countries, and large-scale randomized cross-national experiments are available. 

Here we use three LLMs (GPT-4o-mini, Gemini 2.5 Flash Lite, and Claude 3 Haiku) to generate individual-level synthetic responses for 59,508 respondents across 62 countries and 11 climate-psychology interventions in the International Climate Psychology Collaboration (ICPC) dataset \cite{RN28,RN29}. Each synthetic response is conditioned on the profile of a real participant, and analyses run at both the population and individual levels across prompting strategies. We first establish the relationship between statistical realism and treatment-effect accuracy, then trace the sources and the consequences of this relationship. Perturbation probes, with population profiles held fixed, are designed to trace the sources of treatment effect error. Two further cross-national experiments \cite{RN30,RN31}, spanning 12 countries (n = 5,948) and 27 countries (n = 14,837), allow us to test whether the relationship generalizes beyond the initial dataset.

Our results show that statistical realism is a weak proxy for treatment effect accuracy in LLM-generated synthetic behavioral data. Models that match the supervised baselines on statistical realism still exhibit treatment-effect errors approximately twice as large as those baselines. Across countries, the rank correlation between statistical realism and treatment-effect accuracy is near zero. Optimizing prompting strategies according to statistical realism can make treatment effect accuracy even worse. These findings caution against using statistical realism as the sole evidence that treatment-effect estimates are accurate, because it can provide misleading reassurance to researchers and policymakers.

This divergence is driven by mismatches in outcome structure and intervention encoding in the simulation process. Errors in statistical realism and treatment-effect accuracy are dominated by different dimensions, and the pattern replicates across models and datasets. Evaluations confined to statistical realism therefore overlook the dimensions along which treatment-effect errors arise. The divergence is larger for behavioral outcomes than for attitudinal ones. LLMs tighten the attitude-behavior association observed in human data, suggesting that simulated behavioral responses partly extrapolate from attitudinal patterns. Consequently, improving statistical realism does not improve treatment-effect accuracy, and gains in statistical realism occur in different populations from gains in accuracy. Based on these findings, we introduce a diagnostic framework for LLM-generated synthetic data and discuss how treatment-effect validation should proceed under varying availability of experimental benchmarks.


\section*{Results}

\subsection*{Statistical realism does not imply treatment-effect accuracy}

Using the ICPC sample, each LLM predicted climate belief (0-100), policy support (0-100), and pro-environmental action (0-8, converted to 0-100). We operationalized statistical realism primarily as individual-level mean absolute error (MAE) between simulated and observed construct scores. Other metrics, including mean error, country-level Spearman rank correlation, and k-nearest-neighbor overlap provide complementary assessments of aggregate response properties (Supplementary Fig.~\ref{fig:s1}). Treatment-effect accuracy is quantified by the absolute difference between the predicted ATE and the observed ATE. The same pipeline was applied to two additional cross-national experiments: Spampatti et al. \cite{RN30}, with 6 psychological inoculation interventions on climate disinformation across 12 countries, and Većkalov et al. \cite{RN31}, with 2 interventions on scientific-consensus communication in 27 countries.

\begin{figure*}[!t]
  \centering
  \includegraphics[width=0.9\linewidth]{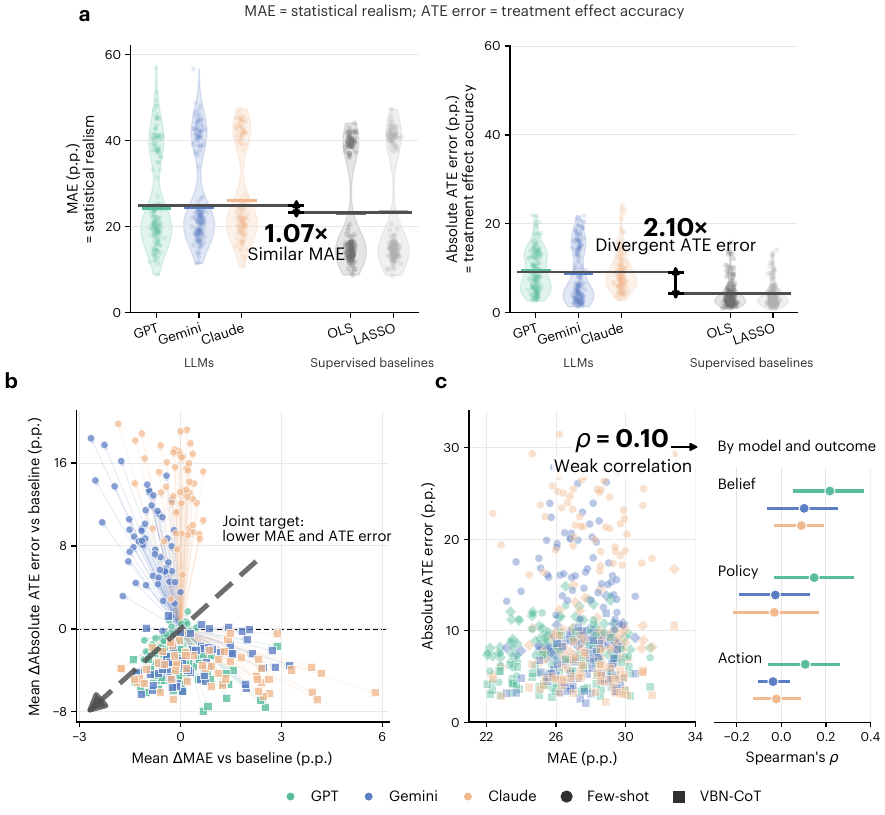}
  \caption{\textbf{Statistical realism does not imply treatment-effect accuracy.} Statistical realism is measured by individual-level mean absolute error (MAE), computed after averaging item-level responses to construct scores. Treatment-effect accuracy is measured by the absolute difference between the predicted and the observed average treatment effect (ATE error). Results are from the ICPC dataset (11 interventions, 62 countries, $n$ = 59,508. \textbf{a}. Similar MAE can coexist with very different ATE accuracy. Comparing the three LLMs with two supervised baselines (OLS and LASSO), MAE is nearly identical (1.07-fold) while absolute ATE error differs more than twofold (2.10-fold). Fold values are the ratio of the mean error across the three LLMs to the mean error across the two supervised baselines, with errors pooled over the three outcomes before the ratio is taken. \textbf{b}. Optimizing MAE does not necessarily improve ATE accuracy. Relative to baseline, reductions in MAE coincide with increases in absolute ATE error, so optimizing statistical realism can move treatment-effect accuracy in the opposite direction. \textbf{c}. MAE and ATE error are only weakly related (Spearman $\rho$ = 0.10).}
  \label{fig:fig2}
\end{figure*}

\textbf{Level: similar statistical realism, different treatment-effect accuracy}. We first established the level of statistical realism in the synthetic data and tested its carry-over to treatment-effect estimation. We compared the LLMs with two supervised baselines, ordinary least squares (OLS) and LASSO regression, under the same outcomes and covariates (Fig.~\ref{fig:fig2}a). On statistical realism, the two sets of models performed similarly. LLM MAEs ranged from 24.3 to 26.1 p.p., close to 23.0 and 23.7 p.p. for OLS and LASSO. The mean MAE across the three LLMs, pooled over the three outcomes, was 1.07 times the mean across the two baselines. This similarity in MAE did not carry over to treatment-effect accuracy. The LLMs missed the observed ATEs by 8.7 to 9.4 p.p., whereas OLS and LASSO missed them by only 4.3 to 4.4 p.p., a 2.10-fold difference in ATE error computed in the same way. Models that look comparable on statistical realism can produce much less accurate treatment-effect estimates.

The outcome breakdown qualifies this contrast in two ways. The mismatch between MAE and ATE error holds within each outcome. In belief, policy support, and action, LLMs were close to OLS and LASSO in MAE, while their ATE errors were more separated. The absolute size of both errors, however, concentrated in action. Belief and policy support had lower MAEs, around 15 to 18 p.p. for the LLMs, and action MAEs rose to 40.7 to 43.2 p.p. ATE error followed the same ordering, with action contributing the largest errors across model classes. Action thus drove much of the overall error level, and similar MAE and different ATE accuracy still coexisted within each outcome type.

\textbf{Optimization: selecting on statistical realism worsens treatment-effect accuracy}. We next examined whether prompts selected by statistical realism would also improve treatment-effect estimates. We used the control group as a development sample (\textit{n} = 5,093) to select two prompting strategies from a broader set of candidates (Supplementary Table~\ref{tab:s2}). KNN-based few-shot prompting (abbr. few-shot) supplied each model with externally retrieved examples from similar respondents, while VBN-based chain-of-thought prompting (abbr. VBN-CoT) asked the model to infer outcomes through the Value-Belief-Norm theory (detailed in Methods). Because the development sample contained no treated participants, prompt selection could rely on individual-level fit but had no access to treatment-effect error, as in applied settings where ground-truth treatment effects have not yet been observed.

The prompt selected by this criterion moved treatment-effect estimates in the wrong direction. Few-shot produced the lowest MAE in all three LLMs, with errors of 22.8 to 25.0 p.p. After applying few-shot, ATE error increased in 8 of the 9 model-outcome cells, with the largest increases for action, from 16.4 to 36.7 p.p. for Gemini and from 8.4 to 44.2 p.p. for Claude. VBN-CoT reduced ATE error in 8 of the 9 cells and produced the lowest overall ATE error in all three LLMs, with errors of 5.4 to 5.8 p.p. Its MAE changes, however, were mixed across models and outcomes, so the MAE criterion did not consistently identify it. Relative to baseline, configurations with larger reductions in MAE tended to show larger increases in absolute ATE error (Spearman $\rho$ = -0.52, p = 0.027, 95\% CI -0.79 to -0.07; descriptive; Fig.~\ref{fig:fig2}b). In this validation setting, optimizing prompts toward individual-level fit would have favored a strategy that more often worsened treatment-effect estimates, while missing the strategy that more often improved them.

\textbf{Ranking: statistical realism does not track treatment-effect accuracy across countries}. Country-level statistical realism shows only a weak relation to treatment-effect accuracy in the ICPC dataset. Pooling country-level observations across models and prompting strategies, the Spearman correlation between MAE and absolute ATE error was 0.10 (\textit{p} = 0.026, bootstrap 95\% CI: 0.01 to 0.19; Fig.~\ref{fig:fig2}c). Across model-outcome combinations, the correlations ranged from -0.04 to 0.22. Most confidence intervals overlapped zero, with only the belief outcome for GPT showing a statistically significant association. Countries with more realistic synthetic outcomes therefore do not consistently yield more accurate treatment-effect estimates.

Country rankings show the same limited alignment. For GPT, the 10 countries with the lowest MAE and the 10 with the lowest ATE error shared only three members (Chile, China, and Portugal), and the 10 highest shared only one (Morocco; Supplementary Fig.~\ref{fig:s5}). Ranking countries by statistical realism therefore gives limited information about where treatment-effect estimates are most accurate or most error-prone.

We further tested the generality of this conclusion in two additional cross-national experiments (Supplementary Fig.~\ref{fig:s2}). Across datasets, statistical realism did not show consistent relationship with treatment-effect accuracy. Models, prompts, and populations that appeared more statistically realistic did not reliably yield more accurate treatment-effect estimates. These replications indicate that response-level realism alone is insufficient for validating simulation-derived treatment effects.

\subsection*{Statistical realism errors and treatment-effect errors are structured differently}

\begin{figure*}[t]
  \centering
  \includegraphics[width=\linewidth]{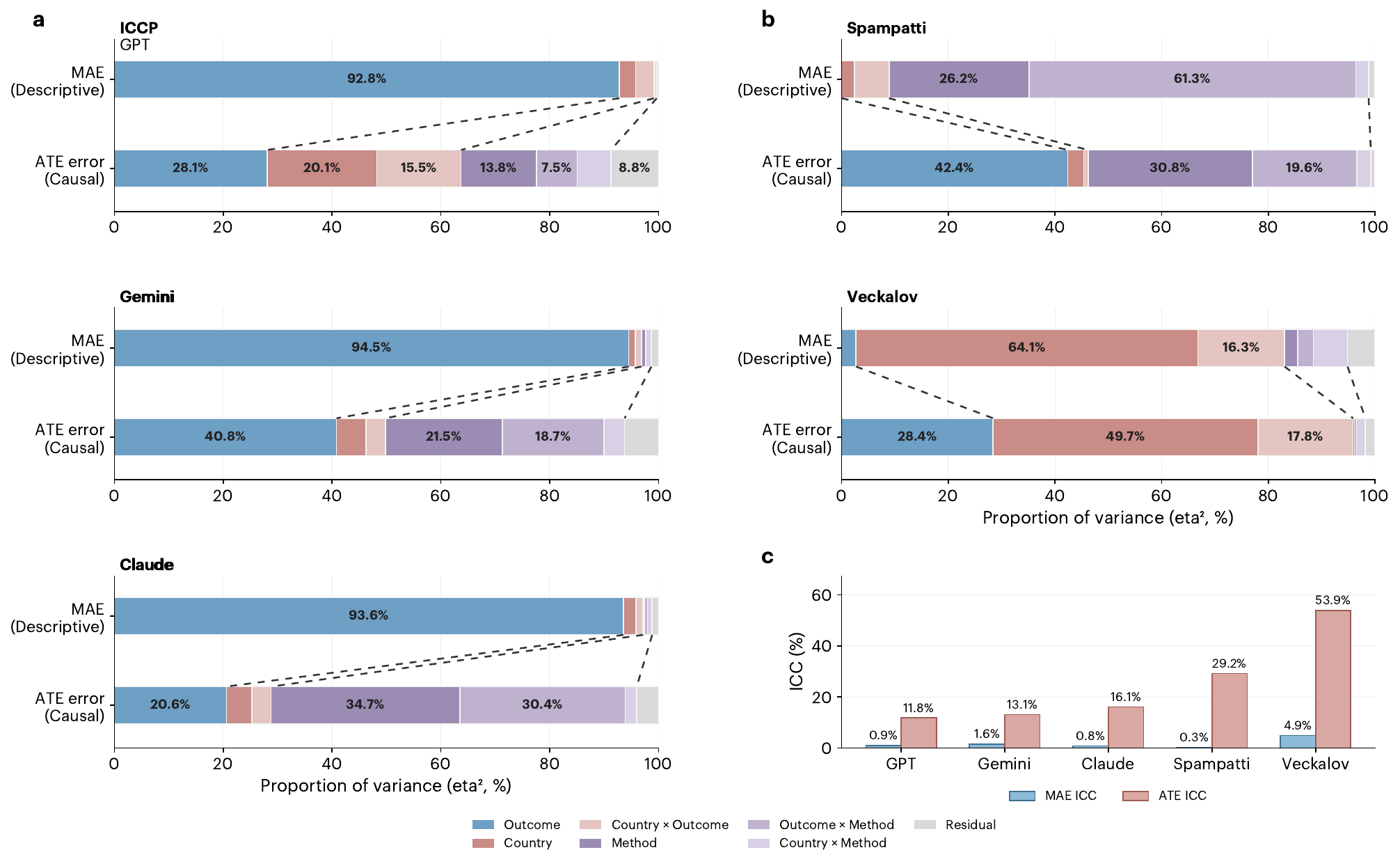}
  \caption{\textbf{Errors of statistical realism and treatment-effect accuracy follow different structures across models and datasets}. \textbf{a}. In ICPC dataset, outcome type dominates MAE, but ATE error redistributes across method, country, and their interactions. \textbf{b}. The structural shift replicates in both additional datasets, although the dimensions involved vary with experimental design. \textbf{c}. Intraclass correlation coefficients (ICC) show that country-level heterogeneity is amplified in treatment-effect accuracy across all model-dataset combinations.}
  \label{fig:fig3}
\end{figure*}

The divergence between statistical realism and treatment-effect accuracy extends beyond their averages to their error structures (Fig.~\ref{fig:fig3}a). For statistical realism, error variance was dominated by outcome type in all three models (92.8\%-94.5\%), and prompting method contributed less than 1\%. The structure of treatment-effect error differed. Outcome type declined to 20.6\%-40.8\%, while method and outcome-by-method interactions jointly accounted for 21.4\%-65.1\% of variance. All three models showed similar qualitative shifts from outcome-dominated error to a more distributed structure, although the dimensions most amplified differed. GPT showed greater country-related heterogeneity (41.8\% from country and its interactions), Claude the strongest amplification of method and its interactions (65.1\% combined), and Gemini retained both a large outcome main effect (40.8\%) and method-related variance (40.2\% combined for method and outcome-by-method interactions). Practically, the factors that drive treatment-effect error contribute little to statistical realism error, so an evaluation confined to realism may look for error in the wrong places.

The two errors were also structured differently in the replication datasets, although the baseline structures themselves differed from ICPC (Fig.~\ref{fig:fig3}b). In the Spampatti dataset, MAE variance was dominated by method and its interaction with country (87.5\% combined), and ATE error variance shifted toward outcome (42.4\%) alongside method (30.8\%). In the Većkalov dataset, MAE variance was dominated by country (64.1\%), and in ATE error the contribution of outcome rose from 2.7\% to 28.4\% while that of country fell to 49.7\%. The dominant dimensions differ with experimental design, but in all three datasets the two errors follow different structures.

Multilevel and permutation analyses further specified this structural difference. Intraclass correlation coefficients (ICC) for country random intercepts increased from 0.8\%-1.6\% for MAE to 11.8\%-16.1\% for absolute ATE error (Fig.~\ref{fig:fig3}c), indicating greater between-country heterogeneity in treatment-effect accuracy. The amplification was larger in the replication datasets (Spampatti: 0.3\% to 29.2\%, Većkalov: 4.9\% to 53.9\%). Permutation tests shuffling responses within country-by-outcome blocks further showed that the structural difference between MAE and ATE error was unlikely to arise from block composition alone in all three datasets (all \textit{p} < 0.001). Statistical realism and treatment-effect accuracy are therefore shaped by different sources of error, and evaluation confined to statistical realism misses the structure that drives treatment-effect accuracy.

\subsection*{Behavioral effect estimates are insensitive to persuasive content of interventions and are overly coupled to attitudes}

\begin{figure*}[!t]
  \centering
  \includegraphics[width=\linewidth]{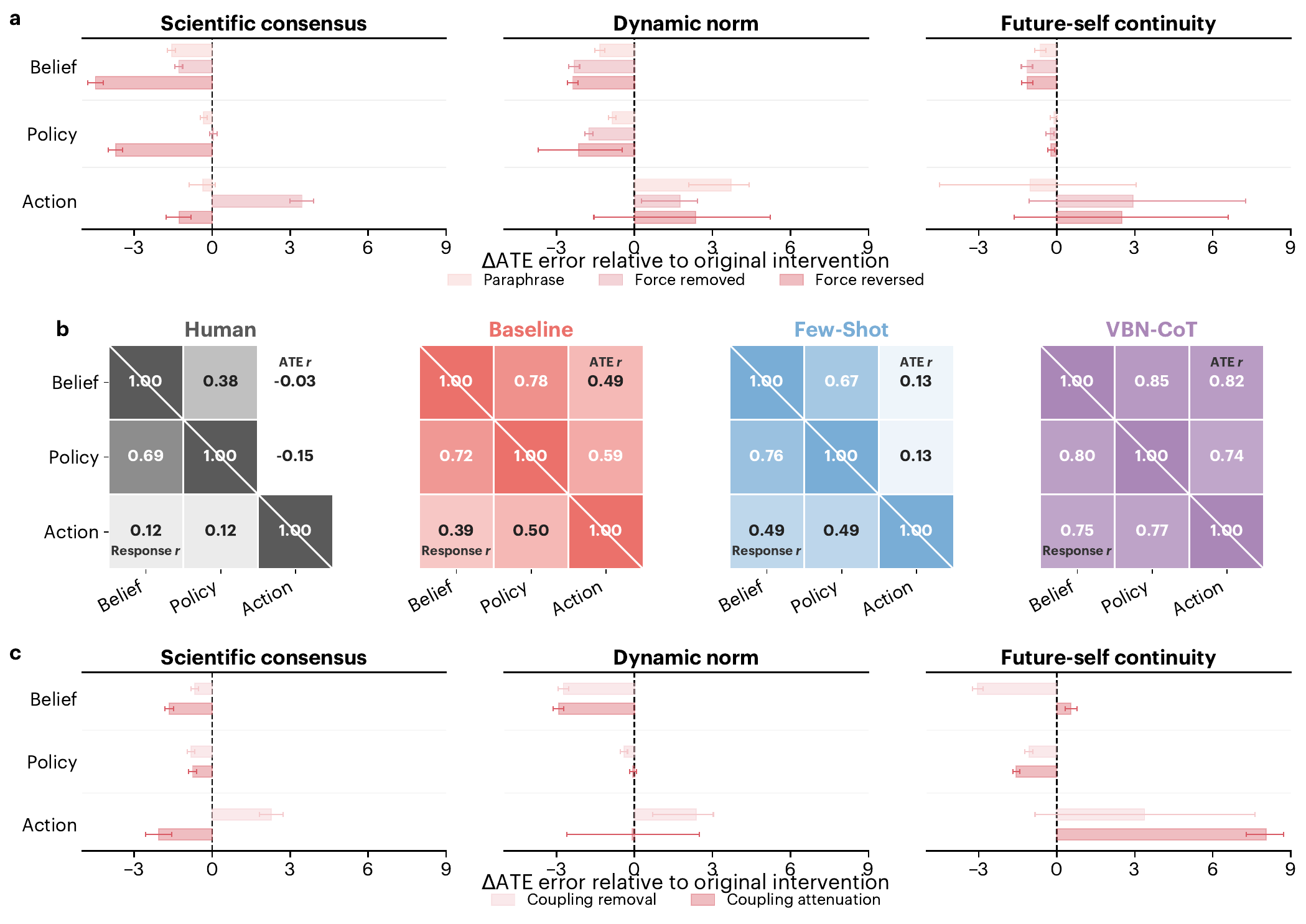}
  \caption{\textbf{Attitudinal estimates track the direction of persuasive force, and models tighten the association between behavioral and attitudinal estimates more than human data do}. Results shown for GPT; other models in Supplementary Figs. ~\ref{fig:s2} and ~\ref{fig:s3}. \textbf{a}. Models are sensitive to persuasive force in intervention text for attitudinal outcomes, but action estimates do not respond coherently to the same perturbations. Bars show the change in absolute ATE error relative to the original intervention text under text perturbations. Error bars indicate 95\% CI. \textbf{b}. LLM-generated data tighten the association between attitudinal and behavioral outcomes relative to human data. Lower triangles show individual-level response correlations. Upper triangles show intervention-level ATE correlations. \textbf{c}. Instructing models to weaken this association does not consistently reduce action ATE error. Corrections improve some interventions but overcorrect others.}
  \label{fig:fig4}
\end{figure*}

We further explored the features of the interventions and outcomes that drive the treatment-effect errors. We designed a set of probes that edited original intervention content to vary persuasive force and preserved overall structure: (1) surface paraphrase, (2) force removal, and (3) force reversal (Supplementary Tables~\ref{tab:s5}, \ref{tab:s6}, and \ref{tab:s7}). As treatment-effect error varied by intervention category (Supplementary Fig.~\ref{fig:s1}), we selected three representative interventions according to their category and range of error magnitudes, including scientific consensus from deductive reasoning, dynamic norm from cultural and group norms, and future-self continuity from situational simulation (Fig.~\ref{fig:fig4}a). Signed ATE bias ($delta$ATE = predicted ATE minus observed ATE) was calculated to distinguish overestimation from underestimation. Compared with the two milder edits, force reversal produced similar or larger reductions in attitudinal ATE overestimation, especially for scientific consensus (belief: -4.49; policy: -3.71) and dynamic norm (belief: -2.37; policy: -2.15). For action, by contrast, changes were mixed in sign and magnitude (e.g., scientific consensus: paraphrase -0.36, force removed +3.46, force reversed -1.27). Attitudinal estimates therefore generally tracked the direction of persuasive force across edits, but behavioral estimates showed no consistent directional pattern.

To understand why behavioral estimates shifted without tracking the direction of persuasive force, we compared the association between attitudinal and behavioral outcomes in the human data and in the model-generated data (Fig.~\ref{fig:fig4}b). In the human data, belief and action are weakly associated at both individual response level ($r_{\mathrm{response}} = 0.11$) and at the intervention level ($r_{\mathrm{ATE}} = -0.03$). Policy and action show the same separation ($r_{\mathrm{response}} = 0.12$, $r_{\mathrm{ATE}} = -0.15$), consistent with the value-action gap (or intention-behavior gap). In LLM-generated data, these associations were stronger, although the three outcomes were elicited separately. For instance, the belief-action $r_{\mathrm{response}}$ rises from 0.11 to 0.39 under baseline, 0.49 under few-shot, and 0.75 under VBN-CoT. The belief-action $r_{\mathrm{ATE}}$ rises from -0.03 to 0.13 under few-shot and 0.82 under VBN-CoT.

Instructing the model to loosen this association did not produce consistent improvement (Fig.~\ref{fig:fig4}c, Supplementary Table~\ref{tab:s8}). We tested both removing the assumed association (coupling removal) and attenuating it by stating that behavioral change is typically smaller and slower than attitudinal change (coupling attenuation). Coupling removal increased action ATE error for scientific consensus by 2.28 p.p., for dynamic norm by 2.37 p.p., and for future-self continuity by 3.38 p.p. Coupling attenuation reduced action ATE error for scientific consensus by 2.06 p.p., produced little change for dynamic norm (-0.08 p.p.), and increased error substantially for future-self continuity by 8.05 p.p., which magnified its underestimation. By contrast, the same manipulations generally reduced ATE error for belief and policy outcomes. For action, the same instruction helped, left unchanged, or worsened the estimate depending on the intervention. The tightened association therefore accounts for only part of the behavioral error. The remaining part arises because the models appear not to capture how specific interventions translate into behavior. Individual-level statistical realism reveals neither part.

\subsection*{Prompting creates unequal improvements in statistical realism but not in treatment-effect accuracy}

\begin{figure*}[!t]
  \centering
  \includegraphics[width=\linewidth]{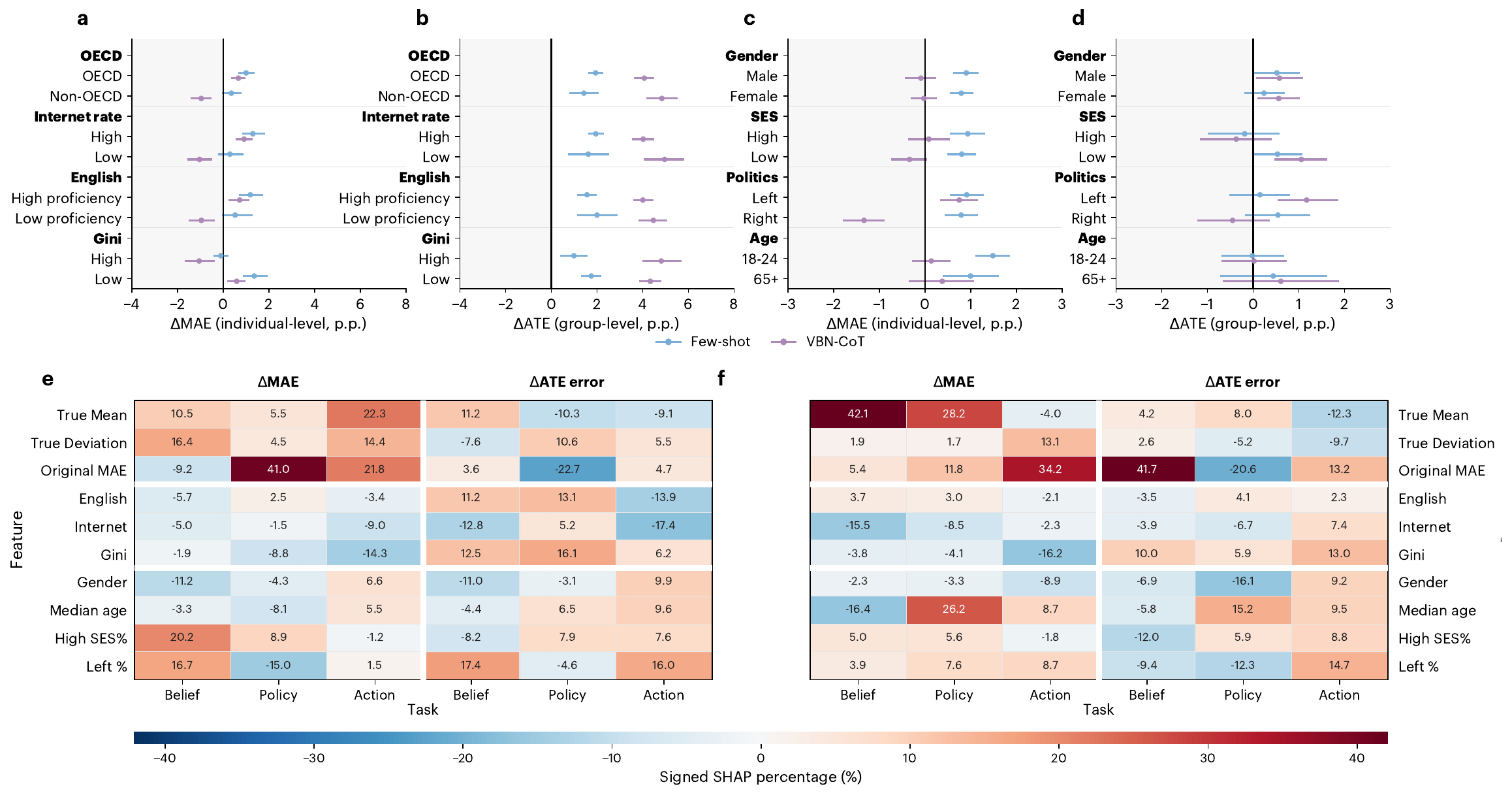}
  \caption{\textbf{Prompting creates unequal gains in statistical realism but not in treatment effect accuracy}. Results shown for GPT; other models in Supplementary Information. \textbf{a-b}. Statistical realism gains concentrated in OECD countries and in countries with higher internet penetration, English proficiency, and lower inequality, but these country-level gradients had no counterpart in treatment effect accuracy gains. Error bars show 95\% CI. \textbf{c-d}. Demographic patterns are weaker and more method-dependent than country-level gradients for both statistical realism and treatment effect accuracy gains. Error bars show 95\% CI. \textbf{e-f}. Feature importance profiles for statistical realism and treatment effect accuracy gains. Statistical realism gains are largely associated with statistical properties (e.g., true mean, original MAE), while treatment effect accuracy gains show a more heterogeneous mix of statistical, demographic, and country-level features. Signed SHAP percentage values from random forest models (100-seed averaged) predicting prompting-induced changes in MAE and ATE error across outcomes. Positive values (red) indicate larger error increases; negative values (blue) indicate larger error reductions. Values are defined relative to the baseline prompt.}
  \label{fig:fig5}
\end{figure*}

To test whether the divergence between statistical realism and treatment-effect accuracy held uniformly across populations, we examined two forms of inequality, baseline inequality in where models are inaccurate and refinement inequality in where prompting improves accuracy. At baseline, socioeconomic characteristics did not consistently predict simulation performance across evaluation targets. Countries associated with lower MAE were not necessarily those with lower ATE error (Supplementary Tables~\ref{tab:s10}). The distribution of prompting gains showed a sharper divergence. At the country level, prompting produced a clear socioeconomic gradient in statistical realism (Fig.~\ref{fig:fig5}a). MAE reductions were concentrated in socioeconomically advantaged countries under both prompting methods. OECD countries, and those with higher internet penetration, higher English proficiency, and lower Gini inequality showed larger improvements (0.6 to 2.0 p.p.; most adjusted \textit{p} < .05 for both methods except English proficiency under few-shot). Reductions in ATE error showed a different pattern (Fig.~\ref{fig:fig5}b), with none of the analyzed country-level characteristics associated with lower ATE error under either prompting strategy (all adjusted \textit{p} > .05). Prompting thus created a broad and method-consistent socioeconomic gradient in statistical realism, whereas corresponding socioeconomic gradients in treatment-effect accuracy were not observed.

Demographic patterns were weaker and more method-dependent for both MAE and ATE error (Fig.~\ref{fig:fig5}c-d). Under few-shot, neither error showed any demographic association, and under VBN-CoT a few characteristics, most notably political orientation, were associated with larger improvements in both errors (all comparisons in Supplementary Tables ~\ref{tab:s11} and~\ref{tab:s12}). Compared with the country level, demographic heterogeneity was more limited, method-specific, and partially overlapping across the two errors.

An exploratory SHAP analysis indicated why the country-level improvements diverged (Fig.~\ref{fig:fig5}e-f). Reductions in MAE were largely associated with the statistical properties of the task. The country’s true response mean and baseline prediction difficulty together dominated the feature profile, with individual contributions reaching 42.1\% (true mean for VBN-CoT belief) and 41.0\% (original MAE for few-shot policy). Reductions in ATE error showed a more heterogeneous mix of statistical, demographic, and country-level features, and the direction of associations shifted across outcome domains and prompting methods. Statistical realism thus improved through relatively predictable task-level features, and treatment-effect accuracy through more heterogeneous, task-specific interactions. Improving or equalizing statistical realism would therefore not necessarily equalize treatment-effect accuracy. The two errors respond to different population-level features. Fairness defined over statistical realism therefore does not transfer to fairness in treatment-effect accuracy. Improving realism for underrepresented groups is a legitimate representational goal, but equalizing the accuracy of treatment effects requires evaluating treatment effects directly.

\section*{Discussion}

Statistical realism has become the default evidence that an LLM simulation can be trusted \cite{RN13}, including for the causal questions such simulations are increasingly used to answer. Whether realism can certify treatment-effect accuracy, however, has not been tested directly. One line of current work supports the feasibility of LLM simulations by demonstrating relatively high fidelity to empirical experimental results \cite{RN20,RN21}, while the other challenges their reliability based on failures in sensitive domains and behavioral decision tasks \cite{RN19}. Despite this divide, both lines of work treat statistical realism as the central criterion for simulation validity, with disagreement focusing on whether models are sufficiently realistic. Recent methodological discussions have further questioned this paradigm by arguing that heuristic validation based on realism cannot guarantee valid extrapolation to causal inference \cite{RN4}. However, this critique remains untested at the empirical level because existing evaluation is conducted along separate dimensions: studies assessing realism typically do not evaluate treatment effects \cite{RN14,RN17}, while studies of behavioral or causal accuracy rarely report statistical realism \cite{RN19,RN32}, preventing the construction of a joint measurement space in which discrepancies between the two could be observed.

We address this gap by jointly evaluating both dimensions on the same set of simulated individuals, thereby constructing the missing joint space and revealing a systematic divergence between statistical realism and treatment-effect accuracy. We tested this link with 11 interventions fielded to 59,508 respondents across 62 countries and replicated the analysis in two additional cross-national datasets. Our results show that the two are weakly related, and improvements in realism can systematically reduce treatment-effect accuracy. Based on our findings, statistical realism alone is insufficient evidence that forward-looking treatment effect estimates (e.g., in generative digital twins and agent societies) can be trusted.

\subsection*{Why statistical realism and treatment effect accuracy diverge}

The divergence arises through two channels, both leaving the statistical structure among responses intact while failing to capture the change an intervention introduces. The first lies in the outcome dimension. Statistical realism rewards a match to the associations among responses, while a treatment effect depends on intervention-induced change \cite{RN33}. In the human data, the association between attitudinal and behavioral outcomes is weak, consistent with the value-action gap \cite{RN34} (or intention-behavior gap \cite{RN26}). The synthetic data tighten the association, and the behavioral estimate tracks the attitudinal ones while it does not respond coherently to the persuasive force of an intervention (Fig.~\ref{fig:fig4}). This tightening is the causal counterpart of a broader tendency for LLM-generated data to inflate associations between variables, reported across tasks \cite{RN14} and in domains such as misinformation sharing \cite{RN35}. Loosening this association improves attitudinal but not behavioral estimates, which suggests the models do not represent how a specific intervention changes behavior. A behavioral response can then stay statistically realistic while its treatment effect follows the tightened association and no longer reflects the change an intervention produces, where statistical realism and treatment-effect accuracy come apart.

The second channel lies in the intervention dimension. LLMs operate on the textual surface of experimental materials, and information that participants draw from lived experience is not in the text. Our probes show that interventions relying on cultural or group norms, which are more explicitly encoded in text, yield more accurate estimates than those requiring individualized or internally simulated experience (e.g., situational simulation; Fig.~\ref{fig:fig4}), consistent with prior findings that text-only LLMs have limited access to embodied or lived experience \cite{RN36,RN37}. When experiential context absent from the text is supplied differently across treatment and control conditions, the comparison underlying the treatment effect is altered, violating identification assumptions such as surrogacy \cite{RN32,RN38} and comparability \cite{RN39}. The statistical surface of the responses can then remain intact while the counterfactual contrast that defines a treatment effect is distorted, leading to the divergence between statistical realism and treatment-effect accuracy. 

\subsection*{Implications for evaluation and application}

Using statistical realism to guide model selection and optimization misleads the decisions it is meant to inform. The mismatch surfaces at three levels, each harder to detect than the last. In model selection and prompting, it misdirects optimization. Optimal settings of statistical realism do not minimize treatment-effect error, and optimizing statistical realism can even degrade it (Fig.~\ref{fig:fig2}), a trade-off also reported in other prompting-optimization tasks \cite{RN40}. At the population level, it undermines fairness interventions, which improve representation for underrepresented groups on the assumption that representational gains carry downstream \cite{RN18}. Higher realism in well-represented populations does not yield higher treatment-effect accuracy (Fig.~\ref{fig:fig5}), so improving representation need not improve fairness in causal estimates. At deployment, the mismatch becomes harder to detect. Applications such as digital twins and policy pretesting aim to predict responses to interventions, a target not constrained by realism, and without ground truth, high realism may become a misleading success signal while errors remain unobserved.

Beyond documenting the divergence, our study contributes a diagnostic framework for evaluating LLM-generated synthetic data against the causal claims they are used to support. The framework jointly evaluates statistical realism and treatment-effect accuracy on the same simulations, localizes their divergence across outcomes, interventions, prompting methods and populations, and uses cross-outcome comparisons and targeted perturbations to probe possible sources of error (see Methods). Because LLM data-generating processes are opaque, whereas formal proxy diagnostics in clinical research \cite{RN41} and econometrics \cite{RN42} typically rely on explicit structural and identification assumptions, these output-based diagnostics can help identify and localize failure modes that cannot be inspected directly. In short, our framework diagnoses where statistical realism-based validation fails but does not itself certify treatment-effect accuracy. What validation should look like instead depends on the inferential claim the synthetic data are meant to support and on the reference data available.

Our results provide little support for treating synthetic responses as substitutes for human experimental data when the goal is to estimate treatment effects. Until treatment-effect accuracy is independently established, estimated effects are better used as hypothesis-generating or screening evidence, with directional and ranking information used to screen candidate designs and guide follow-up human studies. When observed experimental effects are available, model-estimated ATEs should be compared against observed effects in model selection, prompt design, and results interpretation, and can be calibrated when suitable labeled data and valid statistical procedures (e.g., prediction-powered inference) are available. When real effects are unavailable but experiments with similar scenarios or intervention logics exist, calibration against those experiments, or small-scale pilot or vignette studies in real populations where feasible, can serve as alternative validation. When neither is available, probing can help test whether synthetic responses change coherently with intervention content, although coherence alone cannot establish effect accuracy. Among all these validation settings, we recommend reporting results separately by outcome domain and by intervention logic wherever the design allows. At the community level, treatment effect benchmarks, validation datasets, and models trained or fine-tuned on experimental data can help make treatment effect evaluation a routine part of LLM-based synthetic data generation.

\subsection*{Limitations and future work}

This work has several limitations. First, our core findings come from climate attitudes and behavior. Although we conducted our analysis on three different datasets, other domains may differ in the degree of attitude-behavior correlation and have different types of interventions. Whether divergence between statistical realism and treatment effect accuracy takes the same form in other behavioral domains warrants further testing. Second, we used lightweight versions of three LLMs and did not conduct fine-tuning. Future research could explore whether this divergence exists in larger or newer models, as well as models fine-tuned for synthetic behavioral data generation. Third, although our main dataset covers more than 60 countries, the national samples may not be uniformly representative of each country’s full population, so the human data used to evaluate treatment effect estimates may be more reliable in some settings than in others. Fourth, we mainly explore the potential mechanisms of the gap between statistical realism and treatment effect accuracy through external response patterns and do not observe them directly in the models’ internal representations. Future work could introduce more fine-grained probing and examine model parameters and training processes to clarify the mechanisms behind the divergence more explicitly.

\section*{Materials and Methods}

\subsection*{Data}
We used the International Climate Psychology Collaboration (ICPC) dataset, a pre-registered cross-national randomized experiment in which 59,508 participants from 62 countries were randomly assigned to one of 11 behavioral interventions or a no-intervention control \cite{RN28,RN29}. We retained three continuous outcomes for analysis: 1) climate change belief (4 items on a 0-100 scale), 2) policy support (9 items on a 0-100 scale), and 3) pro-environmental action (via the Work for Environmental Protection Task \cite{RN43}, 0-8 trees planted and standardized to 0-100).

We replicated core findings on two additional climate-related experimental datasets: Spampatti et al. \cite{RN30} and Većkalov et al. \cite{RN31}. We retained outcomes that most closely mapped onto the focal domains of the ICPC analysis. In the Spampatti dataset, we used climate belief (9 items on a 1-5 scale) and the same Work for Environmental Protection Task as belief and action outcomes, respectively, all rescaled to 0-100. In the Većkalov dataset, we used perceived scientific consensus (2 items on a 0-100 scale), belief and concern (4 items on a 1-7 scale), and policy support (1 item on a 1-7 scale), all rescaled to 0-100. Full variable lists are provided in Supplementary Table~\ref{tab:s9}.

\subsection*{LLM simulation pipeline}
\textbf{Models and prompting}. From the survey measures, we selected four sets of variables as background information for LLMs: 1) Demographic information, including age, gender, country, and education; 2) Socio-economic status, including a direct material-asset measure based on household appliance ownership and an indirect subjective-status measure based on the MacArthur ladder; 3) Political orientation on social issues and economic issues; and 4) Perceived scientific consensus regarding human-caused climate change. The background information was serialized into narrative prompts to instruct the LLMs. For instance:

\textit{“\#YOUR PROFILE: A 39-year-old man living in the US with an education of college/university education (13-16 years) where the person answered the question 'What is your political orientation for social issues? (0-Extremely liberal/left-wing to 100-Extremely conservative/right-wing)' as '87', the person indicated they own/have access to: Separate room for kitchen, Washing machine, Vacuum cleaner, Freezer/deep freeze, Personal computer, Bathroom, and Television, the person answered the question 'To the best of your knowledge, what percentage of climate scientists have concluded that human-caused climate change is happening? (0-100 Percentage)' as '51'……”}

For participants in intervention conditions, the prompt additionally included the intervention text as presented to the original survey respondent. The 11 interventions span three theoretical categories: 1) deductive reasoning, including collective action framing, scientific consensus communication, and pluralistic ignorance correction; 2) situational simulation techniques, including writing a letter to future generations, writing a letter as the future yourself, psychological distance reduction, and negative emotion induction; and 3) cultural and group norms, including system justification reframing, building moral foundations, dynamic social norms, working-together norms. Explanations of the interventions are provided in Supplementary Table~\ref{tab:s4}, and the psychological rationale of designing these interventions can be found at Vlasceanu et. al \cite{RN28,RN29}. The converted intervention texts in the prompts are provided in Supplementary Information.

These prompts were provided to three widely used LLMs in large-scale simulations (GPT-4o-mini, Gemini 2.5 Flash Lite, and Claude 3 Haiku) through their APIs. Models were instructed to return item-level responses in JSON format. We ran separate simulations for the three outcome domains. All models were queried with \texttt{temperature} = 0.

\textbf{Prompting strategies}. We developed and tested two categories of prompting strategies on a development set drawn from the control group (\textit{n} = 5,093). \textbf{External information} methods augmented prompts with real-world data, including KNN-based few-shot retrieval (demographically similar exemplars within the sample’s country), country-level distributional context (socio-economic, demographic, and climate-related information), and target-variable distributional statistics (national means, variances, and medians). \textbf{Internal reasoning} methods guided the model’s inference process: generic chain-of-thought and Value-Belief-Norm (VBN) theory-based chain-of-thought. Details of these methods and complete prompt templates are provided in Supplementary Text S1-S3.

Because the development set contained only control-group responses, no treatment effects could be computed at the selection stage, and selection rested on statistical realism (MAE) alone, the same constraint that applied work faces when prompts are chosen before effects are realized. On this basis we selected two prompting strategies for full evaluation (Supplementary Table~\ref{tab:s2}). Selection therefore rested on statistical realism alone, the same constraint that applied work faces when prompts are chosen before effects are realized. 1) Few-shot: the prompt was augmented with 6 demographically similar exemplars retrieved via KNN within the same country, experimental condition, and outcome domain as the target case. Each exemplar included its observed response on the corresponding outcome, and the target respondent was excluded from retrieval. 2) VBN-based chain of thought (VBN-CoT): the model first inferred the profile’s environmental values, climate change beliefs, and personal norms from their demographic profile (stage 1), then predicted outcome responses conditioned on this intermediate reasoning (stage 2). This design operationalizes the Value-Belief-Norm theory of environmentalism \cite{RN44}, which posits that pro-environmental behavior arises from a causal chain of values – ecological worldview – awareness of consequences – personal norms – behavior. The baseline condition included the full serialized participant profile, without additional retrieved exemplars or reasoning guidance.

The same prompting templates and prompting strategies were applied to the replication datasets.

\textbf{Model outputs}. Each participant was simulated under all three methods. For each simulated participant, we recorded the predicted response on each outcome. Control and treatment participants were simulated using the same profile template. For treatment participants, the assigned intervention text was appended to the prompt. All models were queried at \texttt{temperature} = 0, so that the decoding is near-deterministic, although the APIs do not guarantee identical outputs across calls. Responses that failed JSON parsing, refused the task or returned API errors were re-queried with the same prompt.

\subsection*{Diagnostic framework}

We organize our analysis as a three-step diagnostic framework. Step 1 uses the same synthetic datasets and experimental benchmarks to jointly evaluate statistical realism and treatment-effect accuracy. This step tests for divergence between the two validation targets. Step 2 localizes any divergence by identifying the dimensions along which it concentrates, across outcomes, interventions, and populations. Step 3 probes possible sources of the divergence through controlled manipulation of model inputs. Cross-outcome coherence compares attitude-behavior correlations in synthetic and human data. Perturbation probes vary intervention content within fixed population profiles and measure the change in effect estimates. Because each step relies only on model outputs and observed reference data, the framework applies to models whose internal representations or training processes are inaccessible.

\subsection*{Evaluation}

\textbf{Statistical realism}. The primary measure of statistical realism is individual-level mean absolute error (MAE). For each participant and construct, we first averaged item-level predictions and observed responses within the construct, then computed the absolute difference between the predicted and observed construct scores. MAE was defined as the mean of these person-level absolute errors within each cell defined by country, outcome, and method, capturing how well the model reproduces individual-level agreement between simulated and observed construct scores.

As complementary country-level measures (Supplementary Fig.~\ref{fig:s1}), we computed: 1) mean error (ME), the signed difference between predicted and observed country-level means for each outcome and method, capturing systematic directional bias in whether the model consistently over- or underestimates a country's average attitude; 2) Spearman rank correlations, which measure whether the model preserves the relative ordering of countries; and 3) KNN overlap (\textit{k} = 5), which quantifies whether the model preserves country similarity structure. For KNN overlap, we constructed each country profile from item-level distributional summaries, identified its \textit{k} nearest countries in the observed and in the model-predicted country-profile space, and computed the overlap between the two neighbor sets.

\textbf{Treatment-effect accuracy}. For each country, outcome, method, and intervention cell, we computed predicted ATEs as the mean difference between simulated treatment and control groups, and compared these against observed experimental ATEs from the ICPC randomized experiment. ATE error was defined as the absolute difference between predicted and observed experimental ATEs (|predicted ATE – true ATE|), capturing the magnitude of treatment-effect estimation error regardless of direction. We use this unsigned metric throughout the main text and figures unless otherwise noted.

\textbf{Conventional statistical models}. We compared LLM performance on both statistical realism and treatment-effect accuracy against conventional statistical models as benchmarks using MAE and absolute ATE error. We fitted linear regression and LASSO ($\alpha$ = 1.0) models for each outcome domain using the same demographic and socioeconomic features provided to the LLMs. Models were trained separately for each experimental condition (including control). Within each condition, data were split 80/20 into training and test sets. We used 5-fold cross-validation on the training split to estimate out-of-sample performance.

\subsection*{Statistical analysis}
All analyses compare two dependent variables: individual-level MAE (statistical realism) and absolute ATE error (treatment-effect accuracy). Because the replication datasets had more limited designs, with fewer countries, interventions, and outcome domains, we restricted replication analyses to variance decomposition, mixed-effects ICC estimation, and permutation tests.

\textbf{Variance decomposition}. We decomposed the variance in MAE and ATE error using factorial ANOVA. For MAE, individual-level data were aggregated to country-outcome-method cells and modeled with country, outcome, method, and all pairwise interactions as fixed effects. For ATE error, we conducted two analyses: 1) a comparable country-aggregated model using the same factor structure, and 2) an intervention-level model additionally including intervention, method-by-intervention, and outcome-by-intervention terms. All models used Type II sums of squares, with $\eta^2$ computed as the factor sum of squares divided by total sum of squares. Here we chose Type II because it tests each factor after accounting for all other factors at the same or lower order, appropriate given our unbalanced design. Type III sums of squares were computed as a robustness check and yielded consistent conclusions.

\textbf{Mixed-effects models}. To account for the clustering by countries, we fitted linear mixed-effects models with country as a random intercept using REML estimation. For MAE, the model included outcome, method, and their interaction as fixed effects. For ATE error, fixed effects included outcome, method, intervention, and all pairwise interactions. We report variance components and intraclass correlation coefficients (ICC, defined as country variance divided by total variance) to quantify the contribution of country-level heterogeneity to each metric.

\textbf{Permutation test}. To validate the method-related variance component without relying on standard distributional assumptions, we conducted a permutation test with 5,000 iterations. In each iteration, method labels were shuffled within country-outcome blocks, the full factorial ANOVA was refitted, and the sum of all method-block sum-of-squares terms (method main effect, outcome-by-method, and country-by-method) was extracted as the test statistic. The permutation p-value was the proportion of permuted method-block sum-of-squares values greater than or equal to the observed value.

\textbf{Proxy validation analysis}. To assess whether statistical realism serves as a proxy for treatment-effect accuracy, we computed Spearman correlations between country-level MAE and ATE error within each outcome-by-method combination. Country-level MAE was defined as the mean individual MAE within a given country, and each correlation was computed across the 62 countries.

\textbf{Cross-outcome coherence}. To quantify whether LLM simulations impose stronger cross-outcome coherence than observed in human data, we computed Pearson correlation matrices among belief, policy, and action using individual-level construct scores for human responses and for each simulation method. We also computed correlations at the ATE level across countries by correlating country-level ATEs across the three outcomes for each intervention.

\subsection*{Perturbation analysis} 
To probe whether LLM treatment-effect accuracy reflects sensitivity to intervention content versus artifacts of attitude-behavior coupling, we conducted two sets of perturbation experiments on three representative interventions, one from each theoretical category: scientific consensus (deductive reasoning), dynamic social norm (cultural and group norms), and future-self continuity (situational simulation). All perturbation analyses were conducted using GPT.

In the first set, we modified intervention texts along a gradient of persuasive force while preserving overall structure. For each intervention, three perturbed versions were generated using GPT-5.4 thinking with explicit editing instructions: (1) surface paraphrase, in which the original text was reworded without altering argumentative content or persuasive force; (2) force removal, in which persuasive elements (e.g., specific evidence, emotional appeals, or normative cues) were removed while retaining the topic and framing; and (3) force reversal, in which the direction of the persuasive argument was reversed (e.g., arguing against pro-environmental action instead of for it). Generated texts were reviewed by the authors to ensure that only the intended dimension was altered. For each perturbation, the full simulation pipeline was re-run under baseline prompting, and the change in absolute ATE error relative to the original intervention text was computed for each outcome domain.

In the second set, we held intervention content constant and varied how the model was instructed to relate attitudinal and behavioral outcomes. Two framing conditions were tested: (1) coupling removal, in which the model was explicitly instructed to infer attitudes and behaviors independently; and (2) coupling attenuation, in which the prompt included an explicit statement that behavioral change is typically smaller and slower than attitudinal change. Changes in absolute ATE error relative to the unmodified prompt were computed for each outcome domain. The complete original and perturbed intervention texts are provided in Supplementary Information.

\subsection*{Inequality and feature importance analysis}

\textbf{Country-level inequality}. To assess whether cross-national reliability gaps change when moving from statistical realism to treatment-effect accuracy, we stratified countries by four structural characteristics: OECD membership (binary, to capture economic development), internet penetration (tercile split, to capture digital access), English proficiency (tercile split, to capture potential linguistic alignment with model training data), and income inequality (Gini coefficient, tercile split, to capture socio-economic conditions). For each stratification, between-group comparisons focused on the extreme groups (e.g., High vs. Low for English proficiency, and top vs. bottom tercile for internet penetration and Gini). We compared between-group differences in mean MAE / mean ATE error and estimated 95\% bootstrap confidence intervals based on 5,000 resamples. We used pairwise Mann-Whitney U tests with Benjamini-Hochberg correction for multiple comparisons to assess subgroup differences. Except for English proficiency scores, which were obtained from the EF English Proficiency Index (EF EPI), all other country-level data were collected from Our World in Data. Because some country-level indicators were unavailable separately for Taiwan, respondents from Taiwan were merged with China for all country-level analyses. 

\textbf{Demographic subgroup analysis}. To assess whether statistical realism and treatment-effect accuracy identify the same groups as relatively well or poorly simulated, we examined simulation accuracy across four demographic contrasts: gender (male vs. female), subjective socio-economic status (high vs. low), political orientation (left vs. right), and age (18-24 vs. 65+). These contrasts represent salient axes of social and political heterogeneity that are also substantively relevant in climate-attitude research. For each contrast, we computed mean MAE and mean ATE error and compared the size of between-group disparities across statistical realism and treatment-effect accuracy using the same bootstrap and significance-testing procedures as in the country-level inequality analysis.

\textbf{Feature importance exploration}. As an exploratory extension, we trained separate Random Forest models to predict prompting-induced changes in MAE and ATE error relative to baseline, and used SHAP values to characterize feature-level correlates of statistical realism and treatment-effect accuracy disparities. Models were fit at the country-by-outcome-by-method level using country-level characteristics (internet penetration, English proficiency, Gini coefficient), demographic composition (median age, gender ratio, proportion high-SES, proportion left-leaning), and statistical properties of the outcome (true mean, true standard deviation, original MAE) as predictors. This set of predictors covered all grouping variables used in the inequality analyses, along with additional statistical properties of each outcome. We then computed SHAP values to quantify each feature’s directional contribution to the predicted change in error relative to baseline, averaging across 100 random seeds to stabilize estimates.

\section*{Data availability}
\footnotesize
The datasets analyzed in this study are publicly available from the original publications cited in the main text. Information on data sources and access is provided in the corresponding articles. All generated synthetic behavioral data are provided on OSF at https://osf.io/jmqpd.

\section*{Code availability}
\footnotesize
The code used for the analyses is available on OSF at https://osf.io/jmqpd.

\section*{Acknowledgements}
\footnotesize
FJ was supported by the Connaught Fund (Grant Number 520245) and Social Sciences and Humanities Research Council (SSHRC) of Canada (Grant Number 215119, 00169), Seed Grant of the International Network of Educational Institutes (Grant Number 522011) and Canada Research Chair program (CRC-2024-00169). ZL was supported by Social Sciences and Humanities Research Council (SSHRC) of Canada (702-2026-1468). The authors would like to thank Zhixun Yan and Yuxuan Yang from the University of Toronto for their research assistance.

\section*{Conflict of interest}
\footnotesize
The authors declare no competing interests.

\footnotesize
\bibliography{main}

@article{RN13,
   author = {Anthis, Jacy Reese and Liu, Ryan and Richardson, Sean M and Kozlowski, Austin C and Koch, Bernard and Evans, James and Brynjolfsson, Erik and Bernstein, Michael},
   title = {Llm social simulations are a promising research method},
   journal = {arXiv preprint arXiv:2504.02234},
   year = {2025},
   type = {Journal Article}
}

@article{RN23,
   author = {Argyle, Lisa P. and Busby, Ethan C. and Fulda, Nancy and Gubler, Joshua R. and Rytting, Christopher and Wingate, David},
   title = {Out of One, Many: Using Language Models to Simulate Human Samples},
   journal = {Political Analysis},
   volume = {31},
   number = {3},
   pages = {337–351},
   ISSN = {1047-1987},
   DOI = {10.1017/pan.2023.2},
   url = {https://www.cambridge.org/core/product/035D7C8A55B237942FB6DBAD7CAA4E49},
   year = {2023},
   type = {Journal Article}
}

@article{RN3,
   author = {Ashokkumar, Ashwini and Hewitt, Luke and Ghezae, Isaias and Willer, Robb},
   title = {Large language models can predict the results of social science experiments},
   journal = {Nature},
   ISSN = {1476-4687},
   DOI = {10.1038/s41586-026-10742-x},
   url = {https://doi.org/10.1038/s41586-026-10742-x},
   year = {2026},
   type = {Journal Article}
}

@article{RN42,
   author = {Athey, Susan and Chetty, Raj and Imbens, Guido W. and Kang, Hyunseung},
   title = {The Surrogate Index: Combining Short-Term Proxies to Estimate Long-Term Treatment Effects More Rapidly and Precisely},
   journal = {The Review of Economic Studies},
   pages = {rdaf087},
   ISSN = {0034-6527},
   DOI = {10.1093/restud/rdaf087},
   url = {https://doi.org/10.1093/restud/rdaf087},
   year = {2025},
   type = {Journal Article}
}

@inproceedings{RN15,
   author = {Cao, Yong and Liu, Haijiang and Arora, Arnav and Augenstein, Isabelle and Röttger, Paul and Hershcovich, Daniel},
   title = {Specializing Large Language Models to Simulate Survey Response Distributions for Global Populations},
   series = {Proceedings of the 2025 Conference of the Nations of the Americas Chapter of the Association for Computational Linguistics: Human Language Technologies (Volume 1: Long Papers)},
   publisher = {Association for Computational Linguistics},
   pages = {3141–3154},
   ISBN = {979-8-89176-189-6},
   DOI = {10.18653/v1/2025.naacl-long.162},
   url = {https://aclanthology.org/2025.naacl-long.162/
https://doi.org/10.18653/v1/2025.naacl-long.162},
   type = {Conference Proceedings}
}

@article{RN35,
   author = {Choi, Eun Cheol and Young, Lindsay and Ferrara, Emilio},
   title = {Overstating Attitudes, Ignoring Networks: LLM Biases in Simulating Misinformation Susceptibility},
   journal = {Proceedings of the International AAAI Conference on Web and Social Media},
   volume = {20},
   pages = {520–547},
   DOI = {10.1609/icwsm.v20i1.42652},
   year = {2026},
   type = {Journal Article}
}

@article{RN27,
   author = {Conner, Mark and Norman, Paul},
   title = {Attitudes, Intentions, and Behavior Change},
   journal = {Annual Review of Psychology},
   volume = {77},
   number = {Volume 77, 2026},
   pages = {311–337},
   ISSN = {1545-2085},
   DOI = {https://doi.org/10.1146/annurev-psych-013125-042110},
   url = {https://www.annualreviews.org/content/journals/10.1146/annurev-psych-013125-042110},
   year = {2026},
   type = {Journal Article}
}

@article{RN21,
   author = {Cui, Ziyan and Li, Ning and Zhou, Huaikang},
   title = {A large-scale replication of scenario-based experiments in psychology and management using large language models},
   journal = {Nature Computational Science},
   volume = {5},
   number = {8},
   pages = {627–634},
   ISSN = {2662-8457},
   DOI = {10.1038/s43588-025-00840-7},
   url = {https://doi.org/10.1038/s43588-025-00840-7},
   year = {2025},
   type = {Journal Article}
}

@article{RN28,
   author = {Doell, Kimberly C. and Todorova, Boryana and Vlasceanu, Madalina and Bak Coleman, Joseph B. and Pronizius, Ekaterina and Schumann, Philipp and Azevedo, Flavio and Patel, Yash and Berkebile-Wineberg, Michael M. and Brick, Cameron and Lange, Florian and Grayson, Samantha J. and Pei, Yifei and Chakroff, Alek and van den Broek, Karlijn L. and Lamm, Claus and Vlasceanu, Denisa and Constantino, Sara M. and Rathje, Steve and Goldwert, Danielle and Fang, Ke and Aglioti, Salvatore Maria and Alfano, Mark and Alvarado-Yepez, Andy J. and Andersen, Angélica and Anseel, Frederik and Apps, Matthew A. J. and Asadli, Chillar and Awuor, Fonda Jane and Basaglia, Piero and Bélanger, Jocelyn J. and Berger, Sebastian and Bertin, Paul and Białek, Michał and Bialobrzeska, Olga and Blaya-Burgo, Michelle and Bleize, Daniëlle N. M. and Bø, Simen and Boecker, Lea and Boggio, Paulo S. and Borau, Sylvie and Borau, Sylvie and Bos, Björn and Bouguettaya, Ayoub and Brauer, Markus and Brik, Tymofii and Briker, Roman and Brosch, Tobias and Buchel, Ondrej and Buonauro, Daniel and Butalia, Radhika and Carvacho, Héctor and Chamberlain, Sarah A. E. and Chan, Hang-Yee and Chow, Dawn and Chung, Dongil and Cian, Luca and Cohen-Eick, Noa and Contreras-Huerta, Luis Sebastian and Contu, Davide and Cristea, Vladimir and Cutler, Jo and D’Ottone, Silvana and De keersmaecker, Jonas and Delcourt, Sarah and Delouvée, Sylvain and Diel, Kathi and Douglas, Benjamin D. and Drupp, Moritz A. and Dubey, Shreya and Ekmanis, Jānis and Elbaek, Christian T. and Elsherif, Mahmoud and Engelhard, Iris M. and Escher, Yannik A. and Etienne, Tom W. and Farage, Laura and Farias, Ana Rita and Feuerriegel, Stefan and Findor, Andrej and Freira, Lucia and Friese, Malte and Gains, Neil Philip and Gallyamova, Albina and Geiger, Sandra J. and Genschow, Oliver and Gjoneska, Biljana and Gkinopoulos, Theofilos and Goldberg, Beth and Goldenberg, Amit and Gradidge, Sarah and Grassini, Simone and Gray, Kurt and Grelle, Sonja and Griffin, Siobhán M. and Grigoryan, Lusine and Grigoryan, Ani and Grigoryev, Dmitry and Gruber, June and Guilaran, Johnrev and others },
   title = {The International Climate Psychology Collaboration: Climate change-related data collected from 63 countries},
   journal = {Scientific Data},
   volume = {11},
   number = {1},
   pages = {1066},
   ISSN = {2052-4463},
   DOI = {10.1038/s41597-024-03865-1},
   url = {https://doi.org/10.1038/s41597-024-03865-1},
   year = {2024},
   type = {Journal Article}
}

@article{RN37,
   author = {Fedorenko, Evelina and Piantadosi, Steven T. and Gibson, Edward A. F.},
   title = {Language is primarily a tool for communication rather than thought},
   journal = {Nature},
   volume = {630},
   number = {8017},
   pages = {575–586},
   ISSN = {1476-4687},
   DOI = {10.1038/s41586-024-07522-w},
   url = {https://doi.org/10.1038/s41586-024-07522-w},
   year = {2024},
   type = {Journal Article}
}

@article{RN1,
   author = {Fenichel, E. P. and Castillo-Chavez, C. and Ceddia, M. G. and Chowell, G. and Parra, P. A. and Hickling, G. J. and Holloway, G. and Horan, R. and Morin, B. and Perrings, C. and Springborn, M. and Velazquez, L. and Villalobos, C.},
   title = {Adaptive human behavior in epidemiological models},
   journal = {Proc Natl Acad Sci U S A},
   volume = {108},
   number = {15},
   pages = {6306–11},
   ISSN = {0027-8424 (Print)
0027-8424},
   DOI = {10.1073/pnas.1011250108},
   year = {2011},
   type = {Journal Article}
}

@article{RN25,
   author = {Ferraro, Paul J. and Miranda, Juan José},
   title = {Heterogeneous treatment effects and mechanisms in information-based environmental policies: Evidence from a large-scale field experiment},
   journal = {Resource and Energy Economics},
   volume = {35},
   number = {3},
   pages = {356–379},
   ISSN = {0928-7655},
   DOI = {https://doi.org/10.1016/j.reseneeco.2013.04.001},
   url = {https://www.sciencedirect.com/science/article/pii/S0928765513000225},
   year = {2013},
   type = {Journal Article}
}

@article{RN19,
   author = {Gao, Yuan and Lee, Dokyun and Burtch, Gordon and Fazelpour, Sina},
   title = {Take caution in using LLMs as human surrogates},
   journal = {Proceedings of the National Academy of Sciences},
   volume = {122},
   number = {24},
   pages = {e2501660122},
   DOI = {10.1073/pnas.2501660122},
   url = {https://doi.org/10.1073/pnas.2501660122},
   year = {2025},
   type = {Journal Article}
}

@article{RN38,
   author = {Gui, George and Toubia, Olivier},
   title = {The challenge of using llms to simulate human behavior: A causal inference perspective},
   journal = {arXiv preprint arXiv:2312.15524},
   year = {2023},
   type = {Journal Article}
}

@article{RN9,
   author = {Gupta, Aayush and Sheikh, Farahan Raza},
   title = {LLM-Powered Social Digital Twins: A Framework for Simulating Population Behavioral Response to Policy Interventions},
   journal = {arXiv preprint arXiv:2601.06111},
   year = {2026},
   type = {Journal Article}
}

@article{RN7,
   author = {Hackenburg, Kobi and Tappin, Ben M. and Hewitt, Luke and Saunders, Ed and Black, Sid and Lin, Hause and Fist, Catherine and Margetts, Helen and Rand, David G. and Summerfield, Christopher},
   title = {The levers of political persuasion with conversational artificial intelligence},
   journal = {Science},
   volume = {390},
   number = {6777},
   pages = {eaea3884},
   DOI = {10.1126/science.aea3884},
   url = {https://doi.org/10.1126/science.aea3884},
   year = {2025},
   type = {Journal Article}
}

@article{RN20,
   author = {Hewitt, Luke and Ashokkumar, Ashwini and Ghezae, Isaias and Willer, Robb},
   title = {Predicting results of social science experiments using large language models},
   journal = {Preprint},
   year = {2024},
   type = {Journal Article}
}

@article{RN4,
   author = {Hullman, Jessica and Broska, David and Sun, Huaman and Shaw, Aaron},
   title = {This human study did not involve human subjects: Validating LLM simulations as behavioral evidence},
   journal = {arXiv preprint arXiv:2602.15785},
   year = {2026},
   type = {Journal Article}
}

@misc{RN16,
   author = {Kaiser, Carolin and Kaiser, Jakob and Manewitsch, Vladimir and Rau, Lea and Schallner, Rene},
   title = {Simulating Human Opinions with Large Language Models: Opportunities and Challenges for Personalized Survey Data Modeling},
   publisher = {Association for Computing Machinery},
   pages = {82–86},
   DOI = {10.1145/3708319.3733685},
   url = {https://doi.org/10.1145/3708319.3733685},
   year = {2025},
   type = {Conference Paper}
}

@article{RN5,
   author = {Ke, Luoma and Tong, Song and Cheng, Peng and Peng, Kaiping},
   title = {Exploring the frontiers of LLMs in psychological applications: a comprehensive review},
   journal = {Artificial Intelligence Review},
   volume = {58},
   number = {10},
   pages = {305},
   ISSN = {1573-7462},
   DOI = {10.1007/s10462-025-11297-5},
   url = {https://doi.org/10.1007/s10462-025-11297-5},
   year = {2025},
   type = {Journal Article}
}

@article{RN34,
   author = {Kollmuss, Anja and Agyeman, Julian},
   title = {Mind the Gap: Why do people act environmentally and what are the barriers to pro-environmental behavior?},
   journal = {Environmental Education Research},
   volume = {8},
   number = {3},
   pages = {239–260},
   ISSN = {1350-4622},
   DOI = {10.1080/13504620220145401},
   url = {https://doi.org/10.1080/13504620220145401},
   year = {2002},
   type = {Journal Article}
}

@article{RN43,
   author = {Lange, Florian and Dewitte, Siegfried},
   title = {The Work for Environmental Protection Task: A consequential web-based procedure for studying pro-environmental behavior},
   journal = {Behavior Research Methods},
   volume = {54},
   number = {1},
   pages = {133–145},
   ISSN = {1554-3528},
   DOI = {10.3758/s13428-021-01617-2},
   url = {https://doi.org/10.3758/s13428-021-01617-2},
   year = {2022},
   type = {Journal Article}
}

@article{RN22,
   author = {Li, Peiyao and Castelo, Noah and Katona, Zsolt and Sarvary, Miklos},
   title = {Frontiers: Determining the Validity of Large Language Models for Automated Perceptual Analysis},
   journal = {Marketing Science},
   volume = {43},
   number = {2},
   pages = {254–266},
   ISSN = {0732-2399},
   DOI = {10.1287/mksc.2023.0454},
   url = {https://doi.org/10.1287/mksc.2023.0454},
   year = {2024},
   type = {Journal Article}
}

@article{RN6,
   author = {Li, Zonghan and Liu, Yi and Wang, Chunyan and Tong, Song and Peng, Kaiping and Ji, Feng},
   title = {Enhancing behavioral nudges with large language model-based iterative personalization: A field experiment on electricity and hot-water conservation},
   journal = {arXiv preprint arXiv:2604.03881},
   year = {2026},
   type = {Journal Article}
}

@book{RN32,
   author = {Lin, Victoria and Yun, Taedong and Matarić, Maja and Canny, John and Gretton, Arthur and D'Amour, Alexander},
   title = {The Illusion of Intervention: Your LLM-Simulated Experiment is an Observational Study},
   DOI = {10.48550/arXiv.2605.20767},
   year = {2026},
   type = {Book}
}

@article{RN33,
   author = {Ludwig, Jens and Mullainathan, Sendhil and Rambachan, Ashesh},
   title = {Large language models: An applied econometric framework},
   journal = {Annual Review of Economics},
   volume = {18},
   ISSN = {1941-1383},
   year = {2024},
   type = {Journal Article}
}

@article{RN8,
   author = {Manivannan, Ajaykumar and Spaiser, Viktoria and Cann, Tristan J. B. and Evans, James and Everall, Jordan P. and Falkenberg, Max and Garcia, David and Guo, Weisi and Herzog, Rico and Otto, Ilona M. and Oswald, Yannick and Pagan, Nicolò and Pellert, Max and Pilgrim, Charlie and Rodriguez-Pardo, Carlos and Sen, Indira and Vezhnevets, Alexander Sasha},
   title = {Generative AI for climate governance and acceptability-constrained policy design},
   journal = {npj Climate Action},
   volume = {5},
   number = {1},
   pages = {37},
   ISSN = {2731-9814},
   DOI = {10.1038/s44168-026-00362-6},
   url = {https://doi.org/10.1038/s44168-026-00362-6},
   year = {2026},
   type = {Journal Article}
}

@article{RN17,
   author = {Pataranutaporn, Pat and Powdthavee, Nattavudh and Archiwaranguprok, Chayapatr and Maes, Pattie},
   title = {Simulating human well-being with large language models: Systematic validation and misestimation across 64,000 individuals from 64 countries},
   journal = {Proceedings of the National Academy of Sciences},
   volume = {122},
   number = {48},
   pages = {e2519394122},
   DOI = {10.1073/pnas.2519394122},
   url = {https://doi.org/10.1073/pnas.2519394122},
   year = {2025},
   type = {Journal Article}
}

@book{RN39,
   author = {Persson, Joel and Schultzberg, Mårten and Ankargren, Sebastian},
   title = {Statistical Foundations of LLM-based A/B Testing: A Surrogacy Framework for Human Causal Inference},
   DOI = {10.48550/arXiv.2606.17165},
   year = {2026},
   type = {Book}
}

@article{RN11,
   author = {Piao, Jinghua and Yan, Yuwei and Zhang, Jun and Li, Nian and Yan, Junbo and Lan, Xiaochong and Lu, Zhihong and Zheng, Zhiheng and Wang, Jing Yi and Zhou, Di},
   title = {Agentsociety: Large-scale simulation of llm-driven generative agents advances understanding of human behaviors and society},
   year = {2025},
   type = {Journal Article}
}

@article{RN41,
   author = {Prentice, Ross L.},
   title = {Surrogate endpoints in clinical trials: Definition and operational criteria},
   journal = {Statistics in Medicine},
   volume = {8},
   number = {4},
   pages = {431–440},
   ISSN = {0277-6715},
   DOI = {https://doi.org/10.1002/sim.4780080407},
   url = {https://doi.org/10.1002/sim.4780080407},
   year = {1989},
   type = {Journal Article}
}

@article{RN2,
   author = {Ruggeri, Kai and Stock, Friederike and Haslam, S. Alexander and Capraro, Valerio and Boggio, Paulo and Ellemers, Naomi and Cichocka, Aleksandra and Douglas, Karen M. and Rand, David G. and van der Linden, Sander and Cikara, Mina and Finkel, Eli J. and Druckman, James N. and Wohl, Michael J. A. and Petty, Richard E. and Tucker, Joshua A. and Shariff, Azim and Gelfand, Michele and Packer, Dominic and Jetten, Jolanda and Van Lange, Paul A. M. and Pennycook, Gordon and Peters, Ellen and Baicker, Katherine and Crum, Alia and Weeden, Kim A. and Napper, Lucy and Tabri, Nassim and Zaki, Jamil and Skitka, Linda and Kitayama, Shinobu and Mobbs, Dean and Sunstein, Cass R. and Ashcroft-Jones, Sarah and Todsen, Anna Louise and Hajian, Ali and Verra, Sanne and Buehler, Vanessa and Friedemann, Maja and Hecht, Marlene and Mobarak, Rayyan S. and Karakasheva, Ralitsa and Tünte, Markus R. and Yeung, Siu Kit and Rosenbaum, R. Shayna and Lep, Žan and Yamada, Yuki and Hudson, Sa-kiera Tiarra Jolynn and Macchia, Lucía and Soboleva, Irina and Dimant, Eugen and Geiger, Sandra J. and Jarke, Hannes and Wingen, Tobias and Berkessel, Jana B. and Mareva, Silvana and McGill, Lucy and Papa, Francesca and Većkalov, Bojana and Afif, Zeina and Buabang, Eike K. and Landman, Marna and Tavera, Felice and Andrews, Jack L. and Bursalıoğlu, Aslı and Zupan, Zorana and Wagner, Lisa and Navajas, Joaquín and Vranka, Marek and Kasdan, David and Chen, Patricia and Hudson, Kathleen R. and Novak, Lindsay M. and Teas, Paul and Rachev, Nikolay R. and Galizzi, Matteo M. and Milkman, Katherine L. and Petrović, Marija and Van Bavel, Jay J. and Willer, Robb},
   title = {A synthesis of evidence for policy from behavioural science during COVID-19},
   journal = {Nature},
   volume = {625},
   number = {7993},
   pages = {134–147},
   ISSN = {1476-4687},
   DOI = {10.1038/s41586-023-06840-9},
   url = {https://doi.org/10.1038/s41586-023-06840-9},
   year = {2024},
   type = {Journal Article}
}

@inproceedings{RN40,
   author = {Sclar, Melanie and Choi, Yejin and Tsvetkov, Yulia and Suhr, Alane},
   title = {Quantifying Language Models' Sensitivity to Spurious Features in Prompt Design or: How I learned to start worrying about prompt formatting},
   booktitle = {International Conference on Learning Representations},
   volume = {2024},
   pages = {25055–25083},
   type = {Conference Proceedings}
}

@article{RN26,
   author = {Sheeran, Paschal and Webb, Thomas L.},
   title = {The Intention–Behavior Gap},
   journal = {Social and Personality Psychology Compass},
   volume = {10},
   number = {9},
   pages = {503–518},
   ISSN = {1751-9004},
   DOI = {https://doi.org/10.1111/spc3.12265},
   url = {https://doi.org/10.1111/spc3.12265},
   year = {2016},
   type = {Journal Article}
}

@article{RN30,
   author = {Spampatti, Tobia and Hahnel, Ulf J. J. and Trutnevyte, Evelina and Brosch, Tobias},
   title = {Psychological inoculation strategies to fight climate disinformation across 12 countries},
   journal = {Nature Human Behaviour},
   volume = {8},
   number = {2},
   pages = {380–398},
   ISSN = {2397-3374},
   DOI = {10.1038/s41562-023-01736-0},
   url = {https://doi.org/10.1038/s41562-023-01736-0},
   year = {2024},
   type = {Journal Article}
}

@article{RN44,
   author = {Stern, Paul C. and Dietz, Thomas and Abel, Troy and Guagnano, Gregory A. and Kalof, Linda},
   title = {A Value-Belief-Norm Theory of Support for Social Movements: The Case of Environmentalism},
   journal = {Human Ecology Review},
   volume = {6},
   number = {2},
   pages = {81–97},
   ISSN = {10744827, 22040919},
   url = {http://www.jstor.org/stable/24707060},
   year = {1999},
   type = {Journal Article}
}

@article{RN10,
   author = {Toubia, Olivier and Gui, George Z. and Peng, Tianyi and Merlau, Daniel J. and Li, Ang and Chen, Haozhe},
   title = {Database Report: Twin-2K-500: A Data Set for Building Digital Twins of over 2,000 People Based on Their Answers to over 500 Questions},
   journal = {Marketing Science},
   volume = {44},
   number = {6},
   pages = {1446–1455},
   ISSN = {0732-2399},
   DOI = {10.1287/mksc.2025.0262},
   url = {https://doi.org/10.1287/mksc.2025.0262},
   year = {2025},
   type = {Journal Article}
}

@article{RN31,
   author = {Većkalov, Bojana and Geiger, Sandra J. and Bartoš, František and White, Mathew P. and Rutjens, Bastiaan T. and van Harreveld, Frenk and Stablum, Federica and Akın, Berkan and Aldoh, Alaa and Bai, Jinhao and Berglund, Frida and Bratina Zimic, Aleša and Broyles, Margaret and Catania, Andrea and Chen, Airu and Chorzępa, Magdalena and Farahat, Eman and Götz, Jakob and Hoter-Ishay, Bat and Jordan, Gesine and Joustra, Siri and Klingebiel, Jonas and Krajnc, Živa and Krug, Antonia and Andersen, Thomas Lind and Löloff, Johanna and Natarajan, Divya and Newman-Oktan, Sasha and Niehoff, Elena and Paerels, Celeste and Papirmeister, Rachel and Peregrina, Steven and Pohl, Felicia and Remsö, Amanda and Roh, Abigail and Rusyidi, Binahayati and Schmidt, Justus and Shavgulidze, Mariam and Vellinho Nardin, Valentina and Wang, Ruixiang and Warner, Kelly and Wattier, Miranda and Wong, Chloe Y. and Younssi, Mariem and Ruggeri, Kai and van der Linden, Sander},
   title = {A 27-country test of communicating the scientific consensus on climate change},
   journal = {Nature Human Behaviour},
   volume = {8},
   number = {10},
   pages = {1892–1905},
   ISSN = {2397-3374},
   DOI = {10.1038/s41562-024-01928-2},
   url = {https://doi.org/10.1038/s41562-024-01928-2},
   year = {2024},
   type = {Journal Article}
}

@article{RN29,
   author = {Vlasceanu, Madalina and Doell, Kimberly C. and Bak-Coleman, Joseph B. and Todorova, Boryana and Berkebile-Weinberg, Michael M. and Grayson, Samantha J. and Patel, Yash and Goldwert, Danielle and Pei, Yifei and Chakroff, Alek and Pronizius, Ekaterina and van den Broek, Karlijn L. and Vlasceanu, Denisa and Constantino, Sara and Morais, Michael J. and Schumann, Philipp and Rathje, Steve and Fang, Ke and Aglioti, Salvatore Maria and Alfano, Mark and Alvarado-Yepez, Andy J. and Andersen, Angélica and Anseel, Frederik and Apps, Matthew A. J. and Asadli, Chillar and Awuor, Fonda Jane and Azevedo, Flavio and Basaglia, Piero and Bélanger, Jocelyn J. and Berger, Sebastian and Bertin, Paul and Białek, Michał and Bialobrzeska, Olga and Blaya-Burgo, Michelle and Bleize, Daniëlle N. M. and Bø, Simen and Boecker, Lea and Boggio, Paulo S. and Borau, Sylvie and Bos, Björn and Bouguettaya, Ayoub and Brauer, Markus and Brick, Cameron and Brik, Tymofii and Briker, Roman and Brosch, Tobias and Buchel, Ondrej and Buonauro, Daniel and Butalia, Radhika and Carvacho, Héctor and Chamberlain, Sarah A. E. and Chan, Hang-Yee and Chow, Dawn and Chung, Dongil and Cian, Luca and Cohen-Eick, Noa and Contreras-Huerta, Luis Sebastian and Contu, Davide and Cristea, Vladimir and Cutler, Jo and D'Ottone, Silvana and De Keersmaecker, Jonas and Delcourt, Sarah and Delouvée, Sylvain and Diel, Kathi and Douglas, Benjamin D. and Drupp, Moritz A. and Dubey, Shreya and Ekmanis, Jānis and Elbaek, Christian T. and Elsherif, Mahmoud and Engelhard, Iris M. and Escher, Yannik A. and Etienne, Tom W. and Farage, Laura and Farias, Ana Rita and Feuerriegel, Stefan and Findor, Andrej and Freira, Lucia and Friese, Malte and Gains, Neil Philip and Gallyamova, Albina and Geiger, Sandra J. and Genschow, Oliver and Gjoneska, Biljana and Gkinopoulos, Theofilos and Goldberg, Beth and Goldenberg, Amit and Gradidge, Sarah and Grassini, Simone and Gray, Kurt and Grelle, Sonja and Griffin, Siobhán M. and Grigoryan, Lusine and Grigoryan, Ani and Grigoryev, Dmitry and Gruber, June and Guilaran, Johnrev and Hadar, Britt and Hahnel, Ulf J. J. and others },
   title = {Addressing climate change with behavioral science: A global intervention tournament in 63 countries},
   journal = {Science Advances},
   volume = {10},
   number = {6},
   pages = {eadj5778},
   DOI = {10.1126/sciadv.adj5778},
   url = {https://doi.org/10.1126/sciadv.adj5778},
   year = {2024},
   type = {Journal Article}
}

@article{RN18,
   author = {Wang, Angelina and Morgenstern, Jamie and Dickerson, John P.},
   title = {Large language models that replace human participants can harmfully misportray and flatten identity groups},
   journal = {Nature Machine Intelligence},
   volume = {7},
   number = {3},
   pages = {400–411},
   ISSN = {2522-5839},
   DOI = {10.1038/s42256-025-00986-z},
   url = {https://doi.org/10.1038/s42256-025-00986-z},
   year = {2025},
   type = {Journal Article}
}

@article{RN24,
   author = {Xie, Yu},
   title = {Population heterogeneity and causal inference},
   journal = {Proceedings of the National Academy of Sciences},
   volume = {110},
   number = {16},
   pages = {6262–6268},
   DOI = {10.1073/pnas.1303102110},
   url = {https://doi.org/10.1073/pnas.1303102110},
   year = {2013},
   type = {Journal Article}
}

@article{RN14,
   author = {Xie, Yueqi and Liang, Lemeng and Li, Shuzhen and Lu, Yifu and Xiao, Zhiwen and Shi, Mengdi and Huang, Junming and Wang, Mengdi and Xie, Yu},
   title = {Evaluating the statistical realism of LLM-generated social science data},
   journal = {Proceedings of the National Academy of Sciences},
   volume = {123},
   number = {19},
   pages = {e2538145123},
   DOI = {10.1073/pnas.2538145123},
   url = {https://doi.org/10.1073/pnas.2538145123},
   year = {2026},
   type = {Journal Article}
}

@article{RN36,
   author = {Xu, Qihui and Peng, Yingying and Nastase, Samuel A. and Chodorow, Martin and Wu, Minghua and Li, Ping},
   title = {Large language models without grounding recover non-sensorimotor but not sensorimotor features of human concepts},
   journal = {Nature Human Behaviour},
   volume = {9},
   number = {9},
   pages = {1871–1886},
   ISSN = {2397-3374},
   DOI = {10.1038/s41562-025-02203-8},
   url = {https://doi.org/10.1038/s41562-025-02203-8},
   year = {2025},
   type = {Journal Article}
}

@article{RN12,
   author = {Yang, Yitian and Duan, Yiqun and Huang, Linghan and Zhu, Yiqi and Bailo, Francesco and Su, Chunmeizi and Chen, Huaming},
   title = {ScioMind: Cognitively Grounded Multi-Agent Social Simulation with Anchoring-Based Belief Dynamics and Dynamic Profiles},
   journal = {arXiv preprint arXiv:2605.13725},
   year = {2026},
   type = {Journal Article}
}

\onecolumn
\setcounter{table}{0}
\renewcommand{\thetable}{S\arabic{table}}
\setcounter{figure}{0}
\renewcommand{\thefigure}{S\arabic{figure}}

\footnotesize
\section*{Supplementary Information}

\begin{table}[h]
\centering
\caption{\textbf{Individual-level MAE and absolute ATE error of conventional statistical models and LLMs.} All outcomes are reported on a 0--100 scale, and pooled values are equal-weight means of Belief, Policy, and Action. LLM values use baseline zero-shot predictions. OLS and LASSO values use five-fold out-of-fold predictions from demographic models fitted to the pooled participant records across the 12 experimental conditions. Individual MAE is averaged over respondents within outcome. Absolute ATE error is calculated for each country and treatment-versus-control comparison, averaged across treatments within country, and then averaged equally across countries.}
\label{tab:s1}
\begin{tabular}{lccccc}
\toprule
Metric & OLS & LASSO & LLM (GPT) & LLM (Gemini) & LLM (Claude) \\
\midrule
Individual MAE & 22.96 & 23.65 & 24.25 & 24.53 & 26.11 \\
\quad Belief & 15.26 & 15.40 & 15.98 & 16.22 & 17.96 \\
\quad Policy & 13.92 & 14.05 & 16.11 & 15.23 & 17.22 \\
\quad Action & 39.71 & 41.50 & 40.65 & 42.13 & 43.15 \\
\midrule
Absolute ATE error & 4.37 & 4.29 & 9.39 & 8.72 & 9.14 \\
\quad Belief & 3.04 & 3.01 & 7.80 & 6.47 & 12.51 \\
\quad Policy & 2.90 & 2.88 & 5.86 & 3.34 & 6.47 \\
\quad Action & 7.16 & 6.97 & 14.51 & 16.35 & 8.43 \\
\bottomrule
\end{tabular}
\end{table}

\begin{table}[h]
\centering
\caption{\textbf{Development-set performance of five enhancement candidates used for method selection ($n$ = 5,093).} \\ Few-shot supplied responses from demographically similar respondents retrieved within the same country and experimental condition using $K$-nearest neighbor clustering ($k$ = 6). VBN-CoT used a two-stage procedure in which the model first inferred a Value-Belief-Norm profile and then generated survey responses conditioned on that intermediate plan. Country-context and distributional-statistics prompts appended country-level contextual summaries or target-variable distributional benchmarks, respectively. Generic chain-of-thought prompting used a two-stage reasoning plan without VBN-specific structure. Based on development-set performance in individual-level MAE, we selected few-shot and VBN-CoT for full evaluation in the main analyses. Complete implementation details and exact prompt templates are available in the analysis code.}
\label{tab:s2}
\begin{tabular}{lcccccc}
\toprule
 & & \multicolumn{3}{c}{External information} & \multicolumn{2}{c}{Internal inference} \\
\cmidrule(lr){3-5} \cmidrule(lr){6-7}
Metric & Baseline & Few-shot & Country context & Distributional stats & CoT & VBN-CoT \\
\midrule
Individual MAE & 23.81 & 22.86 & 24.09 & 23.67 & 24.91 & 23.76 \\
\quad Belief & 15.79 & 15.80 & 15.59 & 15.90 & 15.59 & 15.68 \\
\quad Policy & 15.41 & 14.84 & 15.19 & 14.56 & 16.41 & 16.39 \\
\quad Action & 40.21 & 37.92 & 41.48 & 40.56 & 42.72 & 39.21 \\
\bottomrule
\end{tabular}
\end{table}

\begin{table}[h]
\centering
\captionsetup[table]{justification=raggedright, singlelinecheck=false}
\caption{\textbf{Prompting-method comparison in the Spampatti and Ve\'{c}kalov replication datasets.}}
\label{tab:s3}
\begin{tabular}{llcccc}
\toprule
Dataset & Method & Individual MAE & MAE rank & Absolute ATE error & ATE-error rank \\
\midrule
Spampatti & Baseline & 28.07 & 2 & 26.52 & 3 \\
Spampatti & Few-shot & 21.09 & 1 & 19.28 & 2 \\
Spampatti & VBN-CoT & 29.70 & 3 & 4.31 & 1 \\
\midrule
Ve\'{c}kalov & Baseline & 10.74 & 2 & 3.20 & 2 \\
Ve\'{c}kalov & Few-shot & 10.73 & 1 & 3.22 & 3 \\
Ve\'{c}kalov & VBN-CoT & 11.51 & 3 & 2.91 & 1 \\
\bottomrule
\end{tabular}
\end{table}

\small
\captionsetup{justification=raggedright, singlelinecheck=false}
\begin{longtable}{l l p{10cm}}
\caption{\textbf{Overview of the 11 climate-psychology interventions used in the ICPC simulations.} \\ Adapted from Figure 1 in Vlasceanu et al. (2024).} \label{tab:s4} \\
\toprule
Category & Intervention & Description \\
\midrule
\endfirsthead
\toprule
Category & Intervention & Description \\
\midrule
\endhead
\midrule
\multicolumn{3}{r}{\textit{Continued on next page}} \\
\endfoot
\bottomrule
\endlastfoot
\multirow{3}{*}{\parbox{2.5cm}{Deductive reasoning}}
  & Collective action & Features examples of successful collective action that have had meaningful effects on climate policies (e.g., protests) or have solved past global issues (e.g., the restoration of the ozone layer). \\\addlinespace[3pt]
  & Scientific consensus & Informs participants that ``99\% of expert climate scientists agree that Earth is warming and climate change is happening, mainly because of human activity.'' \\\addlinespace[3pt]
  & Pluralistic ignorance & Presents real public opinion data collected by the United Nations that show what percentage of people in each participant's country agree that climate change is a global emergency. \\\addlinespace[3pt]
\midrule
\multirow{4}{*}{\parbox{2.5cm}{Situational simulation}}
  & Letter to future generations & Emphasizes how one's current actions affect future generations by asking participants to write a letter to a socially close child who will read it in 25 years when they are an adult, describing current actions toward ensuring a habitable planet. \\\addlinespace[3pt]
  & Future self-continuity & Emphasizes the future self-continuity by asking each participant to project themselves into the future and write a letter addressed to themselves in the present, describing the actions they would have wanted to take regarding climate change. \\\addlinespace[3pt]
  & Psychological distance & Frames climate change as a proximal risk using examples of recent natural disasters caused by climate change in each participant's nation and prompts them to write about the climate impacts on their community. \\\addlinespace[3pt]
  & Negative emotions & Exposes participants to ecologically valid scientific facts regarding the impacts of climate change framed in a ``doom and gloom'' style of messaging that were drawn from different real-world news and media sources. \\\addlinespace[3pt]
\midrule
\multirow{4}{*}{\parbox{2.5cm}{Cultural \& group norms}}
  & System justification & Frames climate change as threatening to the way of life to each participant's nation and makes an appeal to climate action as the patriotic response. \\\addlinespace[3pt]
  & Building moral foundations & Invokes authority (e.g., ``from scientists to experts in the military, there is near universal agreement''), purity (e.g., keep our air, water, and land pure), and ingroup-loyalty (e.g., ``it is the American solution'') moral foundations. \\\addlinespace[3pt]
  & Dynamic social norm & Informs participants of how country-level norms are changing and ``more and more people are becoming concerned about climate change,'' suggesting that people should take action. \\\addlinespace[3pt]
  & Work-together norm & Combines referencing a social norm (i.e., ``a majority of people are taking steps to reduce their carbon footprint'') with an invitation to ``join in'' and work together with fellow citizens toward this common goal. \\
\end{longtable}

\begin{longtable}{p{4cm} p{4cm} p{4cm} p{4cm}}
\caption{\textbf{Perturbed texts for the Scientific Consensus intervention.} \\\uline{\textbf{Underlined bold}} text indicates key modified phrases.} \label{tab:s5} \\
\toprule
Original & Surface paraphrase & Force removal & Force reverse \\
\midrule
\endfirsthead
\toprule
Original & Surface paraphrase & Force removal & Force reverse \\
\midrule
\endhead
Did you know that 99\% of expert climate scientists agree that the Earth is warming and climate change is happening, \textbf{\uline{mainly because of human activity (for example, burning fossil fuels)}}? &
Did you know that 99\% of expert climate scientists agree that global warming is accelerating and the climate is changing, primarily driven by human actions (for example, fossil fuel combustion)? &
Did you know that 99\% of expert climate scientists agree that the Earth is warming and climate change is happening, \textbf{\uline{while its causes are still under study (for example, temperature records)}}? &
Did you know that 99\% of expert climate scientists agree that the Earth is warming and climate change is happening, \textbf{\uline{mainly because of natural cyclical processes (for example, solar cycles)}}? \\
\bottomrule
\end{longtable}

\begin{longtable}{p{4cm} p{4cm} p{4cm} p{4cm}}
\caption{\textbf{Perturbed texts for the Dynamic Social Norm intervention.} \\Omitted passages are identical to the original. \uline{\textbf{Underlined bold}} text indicates key modified phrases.} \label{tab:s6} \\
\toprule
Original & Surface paraphrase & Force removal & Force reverse \\
\midrule
\endfirsthead
\toprule
Original & Surface paraphrase & Force removal & Force reverse \\
\midrule
\endhead
\midrule
\multicolumn{4}{r}{\textit{Continued on next page}} \\
\endfoot
\bottomrule
\endlastfoot
People in \{country\} and around the world are changing: more and more people are concerned about climate change, and are now taking action across multiple fronts. Since 2013, concerns about climate change have \textbf{\uline{increased}} in most countries surveyed. &
People in \{country\} and around the world are shifting: a growing number of people are worried about climate change, and are now acting on multiple fronts. Since 2013, worries about climate change have \textbf{\uline{risen}} in most countries surveyed. &
People in \{country\} and around the world hold a range of views on climate change, and show different patterns of engagement across multiple fronts. Since 2013, views on climate change have \textbf{\uline{been recorded}} across most countries surveyed. &
People in \{country\} and around the world are changing: fewer and fewer people are concerned about climate change, and fewer are taking action across multiple fronts. Since 2013, concerns about climate change have \textbf{\uline{declined}} in most countries surveyed. \\
\addlinespace[6pt]
\textit{[Chart: Pew Research Center, Spring 2013 \& 2018 Global Attitudes Surveys. For each country, paired data points connected by a line show that \uline{concern has grown} over the period.]} &
\textit{[Chart: Pew Research Center, Spring 2013 \& 2018 Global Attitudes Surveys. For each country, paired data points connected by a line show that \uline{worry has risen} over the period.]} &
\textit{[Chart: Pew Research Center, Spring 2013 \& 2018 Global Attitudes Surveys. For each country, paired data points connected by a line show \uline{how views have been tracked} over the period.]} &
\textit{[Chart: Pew Research Center, Spring 2013 \& 2018 Global Attitudes Surveys. For each country, paired data points connected by a line show that \uline{concern has declined} over the period.]} \\
\addlinespace[6pt]
What kinds of actions are people taking right now? More than ever before, people are \textbf{\uline{making changes to}} their lifestyles, supporting policies to address climate change, and are giving the issue more time and attention. For example, \textbf{\uline{more and more people from around the world}} are now: [Practice unchanged] &
What kinds of actions are people taking right now? More than before, people are \textbf{\uline{adjusting}} their lifestyles, backing policies to address climate change, and are devoting more time and attention to the issue. For example, \textbf{\uline{a rising number of people from around the world}} are now: [Practice unchanged] &
What kinds of patterns can be observed? People \textbf{\uline{show different patterns in}} their lifestyles, views on policies related to climate change, and the time and attention they give to the issue. For example, \textbf{\uline{people from around the world show varied patterns in}}: [Practice unchanged] &
What kinds of actions are people taking right now? Compared with before, \textbf{\uline{fewer people are making changes}} to their lifestyles, supporting policies to address climate change, and giving the issue time and attention. For example, \textbf{\uline{fewer and fewer people from around the world}} are now: [Practice unchanged] \\
\addlinespace[6pt]
There's also been \textbf{\uline{a notable increase in support}} for climate change mitigation policy--some of \textbf{\uline{the most popular}} policies include: [Policies unchanged] &
There's also been \textbf{\uline{a notable increase in support}} for climate change mitigation policy--some of \textbf{\uline{the most popular}} policies include: [Policies unchanged] &
Views on climate change mitigation policy also \textbf{\uline{vary across contexts}}--some of \textbf{\uline{the most discussed}} policies include: [Policies unchanged] &
There's also been \textbf{\uline{a notable decline in support}} for climate change mitigation policy--some of \textbf{\uline{the least popular}} policies include: [Policies unchanged] \\
\end{longtable}

\begin{longtable}{p{4cm} p{4cm} p{4cm} p{4cm}}
\caption{\textbf{Perturbed texts for the Future-Self Continuity intervention.} \\Omitted passages are identical to the original. \uline{Underlined bold} text indicates key modified phrases.} \label{tab:s7} \\
\toprule
Original & Surface paraphrase & Force removal & Force reverse \\
\midrule
\endfirsthead
\toprule
Original & Surface paraphrase & Force removal & Force reverse \\
\midrule
\endhead
[Climate change description unchanged] &
[Climate change description unchanged] &
[Climate change description unchanged] &
[Climate change description unchanged] \\
\addlinespace[4pt]
[Climate and human activity unchanged] &
[Climate and human activity unchanged] &
[Climate and human activity unchanged] &
[Climate and human activity unchanged] \\
\addlinespace[6pt]
Please put yourself in the year 2030 - eight years from now. Take a few moments to imagine your life in that future. Imagine how you will look, where you will be, and who you are with. In the year 2030, it will be clear \textbf{\uline{whether keeping climate change under 2$^\circ$C is still possible}}. It will be clear \textbf{\uline{whether the necessary change occurred fast enough}} to match the speed of the changing climate. As the Earth's atmosphere continues to heat up, the effects of climate change will \textbf{\uline{be more apparent}}: the ``highest observed temperature'' records will keep being updated, heatwaves and the draughts will become more common, species will continue to become extinct. &
Now picture yourself in the year 2030 - eight years into the future. Take a few moments to envision your life at that point in time. Think about how you will look, where you will be living, and who will be around you. By 2030, it will be evident \textbf{\uline{whether limiting warming to 2$^\circ$C remains an achievable goal}}. It will be clear \textbf{\uline{whether the necessary changes happened quickly enough}} to keep pace with the changing climate. As temperatures continue to climb, the effects of climate change will \textbf{\uline{become more apparent}}: temperature records will keep being broken, heatwaves and droughts will intensify and spread, and more species will face the threat of extinction. &
Please put yourself in the year 2030 - eight years from now. Take a few moments to imagine your life in that future. Imagine how you will look, where you will be, and who you are with. In the year 2030, it will be clearer \textbf{\uline{how likely keeping climate change under 2$^\circ$C still appears to be}}. It will be clear \textbf{\uline{whether various factors played out as projected in current climate models}}. As the Earth's atmosphere continues to heat up, the long-term effects of climate change will \textbf{\uline{become clearer over time}}: the ``highest observed temperature'' records will keep being updated, heatwaves and the droughts will become more common, species will continue to become extinct. &
Please put yourself in the year 2030 - eight years from now. Take a few moments to imagine your life in that future. Imagine how you will look, where you will be, and who you are with. In the year 2030, \textbf{\uline{the trajectory of climate change will have been shaped by forces beyond any individual's control}}. It will be clear \textbf{\uline{whether the changing climate was shaped mainly by forces beyond individual action}}. As the Earth's atmosphere continues to heat up, the effects of climate change will \textbf{\uline{unfold regardless of individual choices}}: the ``highest observed temperature'' records will keep being updated, heatwaves and the droughts will become more common, species will continue to become extinct. \\
\bottomrule
\end{longtable}

\begin{longtable}{p{5.33cm} p{5.33cm} p{5.33cm}}
\caption{\textbf{Prompt framings used in the attitude-behavior coupling probe.} \\\uline{\textbf{Underlined bold}} text indicates the inserted framing statement.} \label{tab:s8} \\
\toprule
Original & Coupling removal & Coupling attenuation \\
\midrule
\endfirsthead
\toprule
Original & Coupling removal & Coupling attenuation \\
\midrule
\endhead
Here is your profile:\newline
\{persona\}\newline
---\newline
\{intervention\}\newline
---\newline
\{question\} &
Here is your profile:\newline
\{persona\}\newline
---\newline
\{intervention\}\newline
\textbf{\uline{Please note, people's views on climate change and their everyday behaviors are related but distinct, and should be inferred separately.}}\newline
---\newline
\{question\} &
Here is your profile:\newline
\{persona\}\newline
---\newline
\{intervention\}\newline
\textbf{\uline{Please note, even when people's views on climate change shift, changes in everyday behavior are often smaller/slower because of habit/cost/social context.}}\newline
---\newline
\{question\} \\
\bottomrule
\end{longtable}

\begin{table}[H]
\centering
\caption{\textbf{Overview of participant profile variables (prompt inputs) and outcome variables across datasets}.\\ All outcomes were rescaled to a common 0--100 metric for cross-study comparison.}
\label{tab:s9}
\small
\begin{tabular}{p{2.2cm} p{7.5cm} p{6.5cm}}
\toprule
\textbf{Dataset} & \textbf{Prompt input variables} & \textbf{Outcome variables} \\
\midrule
ICPC \newline (62 countries, \newline 11 interventions) &
Age, gender, country, education, socioeconomic status (direct material-asset measure and indirect subjective-status measure), political orientation (social and economic issues), perceived scientific consensus &
Climate belief (4 items, 0--100) \newline
Policy support (9 items, 0--100) \newline
Pro-environmental action (WEPT, 0--8 $\rightarrow$ 0--100) \\
\addlinespace
Spampatti et al. \newline (12 countries, \newline 6 interventions) &
Age, gender, country, education, political ideology, cognitive reflection (CRT-2) &
Climate belief (9 items, 1--5 $\rightarrow$ 0--100) \newline
Pro-environmental action (WEPT, 0--8 $\rightarrow$ 0--100) \\
\addlinespace
Ve\'{c}kalov et al. \newline (27 countries, \newline 2 interventions) &
Age, gender, country, urbanicity, student status, college-degree group, political ideology, trust in climate scientists &
Perceived consensus (2 items, 0--100) \newline
Belief and concern (4 items, 1--7 $\rightarrow$ 0--100) \newline
Policy support (1 item, 1--7 $\rightarrow$ 0--100) \\
\bottomrule
\end{tabular}
\end{table}

\begingroup
\setlength{\tabcolsep}{3.0pt}
\renewcommand{\arraystretch}{1.08}
\begin{longtable}{l l *{6}{>{\centering\arraybackslash}p{1.4cm}}}
\caption{\textbf{Pairwise group comparisons under the baseline condition.} Entries report between-group differences (first group minus second group) in individual-level absolute prediction error (MAE) and absolute treatment-effect error (ATE error). All entries are in percentage points and are based on pairwise Mann--Whitney U tests. Significance markers are based on BH-adjusted $p$-values across the eight prespecified contrasts, separately for each model and validation target: $^{***}$ $p_{\mathrm{adj}} < 0.001$, $^{**}$ $p_{\mathrm{adj}} < 0.01$, $^{*}$ $p_{\mathrm{adj}} < 0.05$, $^\dagger$ $p_{\mathrm{adj}} < 0.1$. Unmarked cells were not significant after BH correction.} \label{tab:s10} \\
\toprule
 & & \multicolumn{2}{c}{GPT} & \multicolumn{2}{c}{Gemini} & \multicolumn{2}{c}{Claude} \\
\cmidrule(lr){3-4} \cmidrule(lr){5-6} \cmidrule(lr){7-8}
Group & Contrast & MAE & ATE error & MAE & ATE error & MAE & ATE error \\
\midrule
\endfirsthead
\toprule
 & & \multicolumn{2}{c}{GPT} & \multicolumn{2}{c}{Gemini} & \multicolumn{2}{c}{Claude} \\
\cmidrule(lr){3-4} \cmidrule(lr){5-6} \cmidrule(lr){7-8}
Group & Contrast & MAE & ATE error & MAE & ATE error & MAE & ATE error \\
\midrule
\endhead
OECD & OECD vs.\ Non-OECD & $+1.22^\dagger$ & $-1.25^\dagger$ & $+0.35$ & $+0.13$ & $+0.90^\dagger$ & $-2.26^{***}$ \\
Internet rate & High vs.\ Low & $+1.97^{*}$ & $-1.57^\dagger$ & $+0.90$ & $+0.23$ & $+1.80^{**}$ & $-2.43^{**}$ \\
English proficiency & High vs.\ Low & $+1.76^\dagger$ & $-1.44$ & $+1.23^\dagger$ & $+0.85^\dagger$ & $+1.33^\dagger$ & $-1.76^{*}$ \\
Gini & High vs.\ Low & $-2.29^{**}$ & $+0.87$ & $-0.84^\dagger$ & $+0.10$ & $-0.53$ & $+1.09^\dagger$ \\
Gender & Male vs.\ Female & $+2.53^{***}$ & $+2.12^{***}$ & $+2.29^{***}$ & $+1.76^{**}$ & $+1.45^{***}$ & $+1.83^{***}$ \\
SES & High vs.\ Low & $-1.18^{***}$ & $-0.43$ & $-1.37^{***}$ & $+1.62^{*}$ & $-1.23^{***}$ & $+3.85^{***}$ \\
Political orientation & Left vs.\ Right & $-3.18^{***}$ & $-1.91^\dagger$ & $-1.69^{***}$ & $-1.16^\dagger$ & $-5.27^{***}$ & $-9.39^{***}$ \\
Age & 18--24 vs.\ 65+ & $+2.29^{***}$ & $-2.12$ & $+1.62^{***}$ & $-1.16$ & $-2.69^{***}$ & $-0.52$ \\
\bottomrule
\end{longtable}
\endgroup

\begingroup
\setlength{\tabcolsep}{3.0pt}
\renewcommand{\arraystretch}{1.08}
\begin{longtable}{l l *{6}{>{\centering\arraybackslash}p{1.4cm}}}
\caption{\textbf{Pairwise group comparisons for reductions in MAE.} Entries report between-group differences (first group minus second group) in reductions in individual-level absolute prediction error relative to baseline. All entries are in percentage points and are based on pairwise Mann--Whitney U tests. Significance markers are based on BH-adjusted $p$-values across the eight prespecified contrasts, separately for each model and prompting method: $^{***}$ $p_{\mathrm{adj}} < 0.001$, $^{**}$ $p_{\mathrm{adj}} < 0.01$, $^{*}$ $p_{\mathrm{adj}} < 0.05$, $^\dagger$ $p_{\mathrm{adj}} < 0.1$. Unmarked cells were not significant after BH correction.} \label{tab:s11} \\
\toprule
 & & \multicolumn{2}{c}{GPT} & \multicolumn{2}{c}{Gemini} & \multicolumn{2}{c}{Claude} \\
\cmidrule(lr){3-4} \cmidrule(lr){5-6} \cmidrule(lr){7-8}
Group & Contrast & Few-shot & VBN-CoT & Few-shot & VBN-CoT & Few-shot & VBN-CoT \\
\midrule
\endfirsthead
\toprule
 & & \multicolumn{2}{c}{GPT} & \multicolumn{2}{c}{Gemini} & \multicolumn{2}{c}{Claude} \\
\cmidrule(lr){3-4} \cmidrule(lr){5-6} \cmidrule(lr){7-8}
Group & Contrast & Few-shot & VBN-CoT & Few-shot & VBN-CoT & Few-shot & VBN-CoT \\
\midrule
\endhead
OECD & OECD vs.\ Non-OECD & $+0.64^{*}$ & $+1.62^{***}$ & $-0.12$ & $+1.51^{*}$ & $+0.54$ & $+2.27^{***}$ \\
Internet rate & High vs.\ Low & $+1.00^{**}$ & $+1.95^{***}$ & $+0.10$ & $+1.53^{*}$ & $+0.87$ & $+2.74^{***}$ \\
English proficiency & High vs.\ Low & $+0.67^\dagger$ & $+1.68^{***}$ & $+0.22$ & $+2.10^{*}$ & $+0.32$ & $+2.48^{***}$ \\
Gini & High vs.\ Low & $-1.46^{**}$ & $-1.64^{**}$ & $-0.18$ & $-2.82^{**}$ & $+0.05$ & $-2.38^{***}$ \\
Gender & Male vs.\ Female & $+0.11$ & $-0.07$ & $-0.60^{*}$ & $+1.67^{***}$ & $-0.98^{***}$ & $+0.63^{***}$ \\
SES & High vs.\ Low & $+0.13$ & $+0.42^\dagger$ & $-0.67^{**}$ & $+0.80$ & $-0.73^{**}$ & $+0.92^{***}$ \\
Political orientation & Left vs.\ Right & $+0.12$ & $+2.08^{***}$ & $+0.93^{**}$ & $+2.09^{***}$ & $+0.36$ & $+1.57^{***}$ \\
Age & 18--24 vs.\ 65+ & $+0.49$ & $-0.24^{*}$ & $-2.62^{***}$ & $+6.66^{***}$ & $-5.03^{***}$ & $+0.58^{*}$ \\
\bottomrule
\end{longtable}
\endgroup

\begingroup
\setlength{\tabcolsep}{3.0pt}
\renewcommand{\arraystretch}{1.08}
\begin{longtable}{l l *{6}{>{\centering\arraybackslash}p{1.4cm}}}
\caption{\textbf{Pairwise group comparisons for reductions in ATE error.} Entries report between-group differences (first group minus second group) in reductions in absolute treatment-effect error relative to baseline. All entries are in percentage points and are based on pairwise Mann--Whitney U tests. Significance markers are based on BH-adjusted $p$-values across the eight prespecified contrasts, separately for each model and prompting method: $^{***}$ $p_{\mathrm{adj}} < 0.001$, $^{**}$ $p_{\mathrm{adj}} < 0.01$, $^{*}$ $p_{\mathrm{adj}} < 0.05$, $^\dagger$ $p_{\mathrm{adj}} < 0.1$. Unmarked cells were not significant after BH correction.} \label{tab:s12} \\
\toprule
 & & \multicolumn{2}{c}{GPT} & \multicolumn{2}{c}{Gemini} & \multicolumn{2}{c}{Claude} \\
\cmidrule(lr){3-4} \cmidrule(lr){5-6} \cmidrule(lr){7-8}
Group & Contrast & Few-shot & VBN-CoT & Few-shot & VBN-CoT & Few-shot & VBN-CoT \\
\midrule
\endfirsthead
\toprule
 & & \multicolumn{2}{c}{GPT} & \multicolumn{2}{c}{Gemini} & \multicolumn{2}{c}{Claude} \\
\cmidrule(lr){3-4} \cmidrule(lr){5-6} \cmidrule(lr){7-8}
Group & Contrast & Few-shot & VBN-CoT & Few-shot & VBN-CoT & Few-shot & VBN-CoT \\
\midrule
\endhead
OECD & OECD vs.\ Non-OECD & $+0.51$ & $-0.77$ & $+3.22^{*}$ & $+1.37^{*}$ & $+0.92$ & $-1.27^{*}$ \\
Internet rate & High vs.\ Low & $+0.33$ & $-0.94$ & $+3.14^{*}$ & $+1.66^{*}$ & $+1.33$ & $-1.36^{*}$ \\
English proficiency & High vs.\ Low & $-0.44$ & $-0.46$ & $+4.77^{*}$ & $+2.15^{*}$ & $+1.07$ & $-1.13^\dagger$ \\
Gini & High vs.\ Low & $-0.76$ & $+0.48$ & $-4.91^{**}$ & $-0.58$ & $-2.96^{*}$ & $+0.76$ \\
Gender & Male vs.\ Female & $+0.28$ & $+0.02$ & $+0.49$ & $+0.14$ & $+1.27^{*}$ & $+0.22$ \\
SES & High vs.\ Low & $-0.72$ & $-1.42^{*}$ & $-0.85^\dagger$ & $-1.04^{*}$ & $-1.64^{**}$ & $-0.01$ \\
Political orientation & Left vs.\ Right & $-0.39$ & $+1.62^{**}$ & $+2.87^{**}$ & $+0.26$ & $-3.97^{***}$ & $-5.51^{***}$ \\
Age & 18--24 vs.\ 65+ & $-0.46$ & $-0.59$ & $+4.85^{***}$ & $-0.93^\dagger$ & $+2.47^{**}$ & $+0.07$ \\
\bottomrule
\end{longtable}
\endgroup

\supptextcaption{Baseline prompt template.}
\begin{promptbox}[System]
You are a survey participant. You will be given a profile that contains factual background and prior questionnaire responses, as well as materials that are part of the survey context. Your goal is to complete the survey in a way that reflects how this specific person would likely think, feel, and respond in this situation, given their background and the materials they are presented with. Prioritize consistency with the person's background and prior responses rather than general norms or idealized answers. Follow the response format EXACTLY as requested (e.g., JSON only). No extra keys, no comments, no markdown, no trailing text. Treat each question as part of the same survey session; keep answers internally consistent with the profile and your previous responses.
\end{promptbox}
\begin{promptbox}[User (belief as an example)]
\texttt{\# YOUR PROFILE}\\
{[PROFILE]}\\[4pt]
\texttt{\# SURVEY QUESTION}\\
How accurate do you think these statements are? Answer on a scale from 0 to 100, where 0 means ``not at all accurate'' and 100 means ``extremely accurate''.\\
Questions:\\
Q1: Taking action to fight climate change is necessary to avoid a global catastrophe.\\
Q2: Human activities are causing climate change.\\
Q3: Climate change poses a serious threat to humanity.\\
Q4: Climate change is a global emergency.\\[4pt]
\texttt{\# RESPONSE FORMAT}\\
Respond in the following JSON format only, with no additional text:\\
\texttt{\{"Q1": <number>, "Q2": <number>, "Q3": <number>, "Q4": <number>\}}
\end{promptbox}

\vspace{18pt}
\supptextcaption{Few-shot prompt template}
\begin{promptbox}[System]
Same as baseline.
\end{promptbox}
\begin{promptbox}[User (belief as an example)]
\texttt{\# YOUR PROFILE}\\
{[PROFILE]}\\[4pt]
\texttt{\# SURVEY QUESTION}\\
How accurate do you think these statements are? Answer on a scale from 0 to 100, where 0 means ``not at all accurate'' and 100 means ``extremely accurate''.\\
Questions:\\
Q1: Taking action to fight climate change is necessary to avoid a global catastrophe.\\
Q2: Human activities are causing climate change.\\
Q3: Climate change poses a serious threat to humanity.\\
Q4: Climate change is a global emergency.\\[4pt]
\texttt{\# RESPONSE FORMAT}\\
Respond in the following JSON format only, with no additional text:\\
\texttt{\{"Q1": <number>, "Q2": <number>, "Q3": <number>, "Q4": <number>\}}\\[4pt]
\texttt{\# REFERENCE}\\
Reference answers from similar participants (same country \& condition):\\
-- Example 1: [FEATURE\_1=value | FEATURE\_2=value | FEATURE\_3=value] -> \texttt{\{"Q1": ..., "Q2": ..., "Q3": ..., "Q4": ...\}}\\
-- Example 2: [FEATURE\_1=value | FEATURE\_2=value | FEATURE\_3=value] -> \texttt{\{"Q1": ..., "Q2": ..., "Q3": ..., "Q4": ...\}}\\
-- ...\\
-- Example 6: [FEATURE\_1=value | FEATURE\_2=value | FEATURE\_3=value] -> \texttt{\{"Q1": ..., "Q2": ..., "Q3": ..., "Q4": ...\}}
\end{promptbox}
\vspace{18pt}

\supptextcaption{VBN-CoT prompt template}
\begin{promptbox}[Stage 1 -- System]
You are an expert of the Value-Belief-Norm (VBN) framework for pro-environmental behavior. You are preparing to answer a survey question as a specific person. You will be given a profile that contains factual background and prior questionnaire responses. Your goal is to infer the person's value orientation (Schwartz self-enhancement $\leftrightarrow$ self-transcendence) and then extract VBN components (Awareness of Consequences -- AC; Ascription of Responsibility -- AR; Personal Norm -- PN) from this profile. Output JSON only. No extra keys, no comments, no markdown, no trailing text.
\end{promptbox}
\begin{promptbox}[Stage 1 -- User (belief as an example)]
\texttt{\# Profile:}\\
{[PROFILE]}\\[4pt]
\texttt{\# Question:}\\
How accurate do you think these statements are? Answer on a scale from 0 to 100, where 0 means ``not at all accurate'' and 100 means ``extremely accurate''.\\
Questions:\\
Q1: Taking action to fight climate change is necessary to avoid a global catastrophe.\\
Q2: Human activities are causing climate change.\\
Q3: Climate change poses a serious threat to humanity.\\
Q4: Climate change is a global emergency.\\
Respond in the following JSON format only, with no additional text: \texttt{\{"Q1": <number>, "Q2": <number>, "Q3": <number>, "Q4": <number>\}}\\[4pt]
Return a VBN plan based on the profile for judging the accuracy of climate-change statements:\\
1) evidence: list 2--4 most relevant explicit profile items (must be explicit; no guessing).\\
2) values: Based on this person's political orientation, social position, and education, infer their dominant value orientation on the Schwartz self-enhancement $\leftrightarrow$ self-transcendence dimension.\\
\quad -- self-enhancement: prioritizes personal status, wealth, and authority\\
\quad -- self-transcendence: prioritizes welfare of others, nature, and equality\\
\quad -- mixed: no clear dominant orientation\\
3) VBN labels (derived from the profile and inferred values; interpreted for belief/accuracy judgments):\\
\quad -- AC: Low | Medium | High. How likely does the profile indicate they have strong perceived severity of climate impacts?\\
\quad -- AR: Low | Medium | High. How likely does the profile suggest they attribute climate change to human activity rather than natural causes?\\
\quad -- PN: Low | Medium | High. How much internal pressure would the profile feel to align with pro-environmental positions?\\
4) synthesis: 2--3 sentences (max 50 words) citing which evidence supports the labels, and 2--3 sentences inferring how these VBN factors drive the final answer.\\[4pt]
Output JSON ONLY:\\
\texttt{\{}\\
\quad\texttt{"evidence": ["...", "..."],}\\
\quad\texttt{"values": "self-enhancement | mixed | self-transcendence",}\\
\quad\texttt{"VBN": \{"AC":"Low|Medium|High","AR":"Low|Medium|High","PN":"Low|Medium|High"\},}\\
\quad\texttt{"synthesis": "..."}\\
\texttt{\}}
\end{promptbox}
\begin{promptbox}[Stage 2 -- System]
Same as baseline.
\end{promptbox}
\begin{promptbox}[Stage 2 -- User (belief as an example)]
\texttt{\# Profile:}\\
Here is your profile: [PROFILE]\\[4pt]
Here is a VBN plan extracted from the profile: \_\_PLAN\_JSON\_\_\\[4pt]
Answer the survey question below by applying the VBN plan.\\
Use the VBN plan as a reference framework, but allow nuanced responses based on the specific question context.\\
Follow the required response format EXACTLY.\\[4pt]
How accurate do you think these statements are? Answer on a scale from 0 to 100, where 0 means ``not at all accurate'' and 100 means ``extremely accurate''.\\
Questions:\\
Q1: Taking action to fight climate change is necessary to avoid a global catastrophe.\\
Q2: Human activities are causing climate change.\\
Q3: Climate change poses a serious threat to humanity.\\
Q4: Climate change is a global emergency.\\[4pt]
Respond in the following JSON format only, with no additional text:\\
\texttt{\{"Q1": <number>, "Q2": <number>, "Q3": <number>, "Q4": <number>\}}
\end{promptbox}

\begin{figure}[!b]
  \centering
  \includegraphics[width=\linewidth]{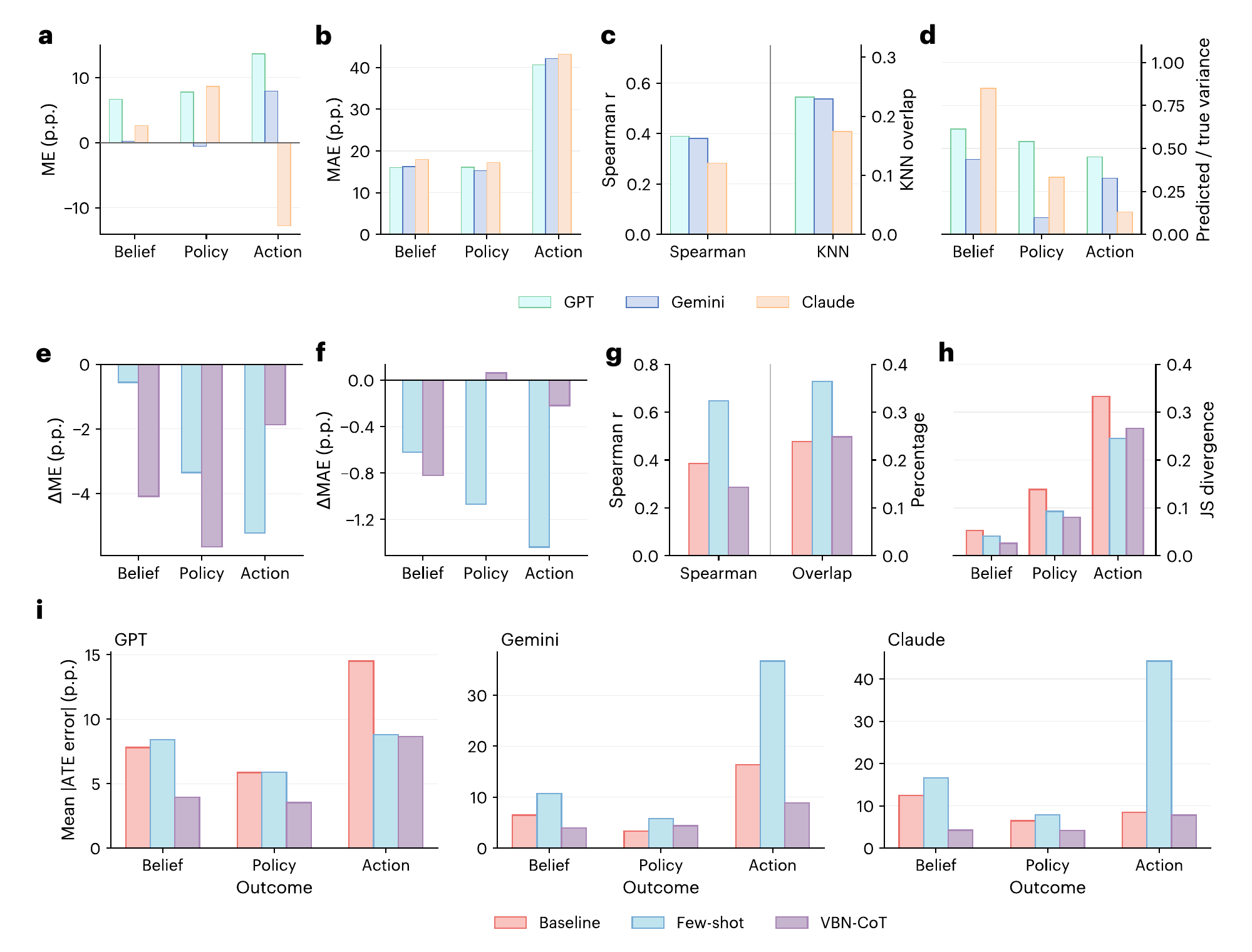}
  \caption{Additional measures of statistical realism and treatment-effect accuracy. \textbf{a}. Mean error (ME) by model and outcome under baseline conditions. \textbf{b}. Mean absolute error (MAE) by model and outcome. \textbf{c}. Country-level Spearman rank correlation and KNN overlap. \textbf{d}. Ratio of predicted to observed variance across models and outcomes. \textbf{e}. Enhancement effect on country-level mean error with prompting strategies. \textit{f}. Enhancement effect on individual-level MAE with prompting strategies. \textbf{g}. Enhancement effect on country-level Spearman correlation and KNN overlap with prompting strategies. \textbf{h}. Enhancement effect on Jensen-Shannon divergence by outcome with prompting strategies. \textbf{i}. Mean absolute ATE error by outcome and prompting strategy for each LLM.}
  \label{fig:s1}
\end{figure}

\begin{figure}
  \centering
  \includegraphics[width=0.9\linewidth]{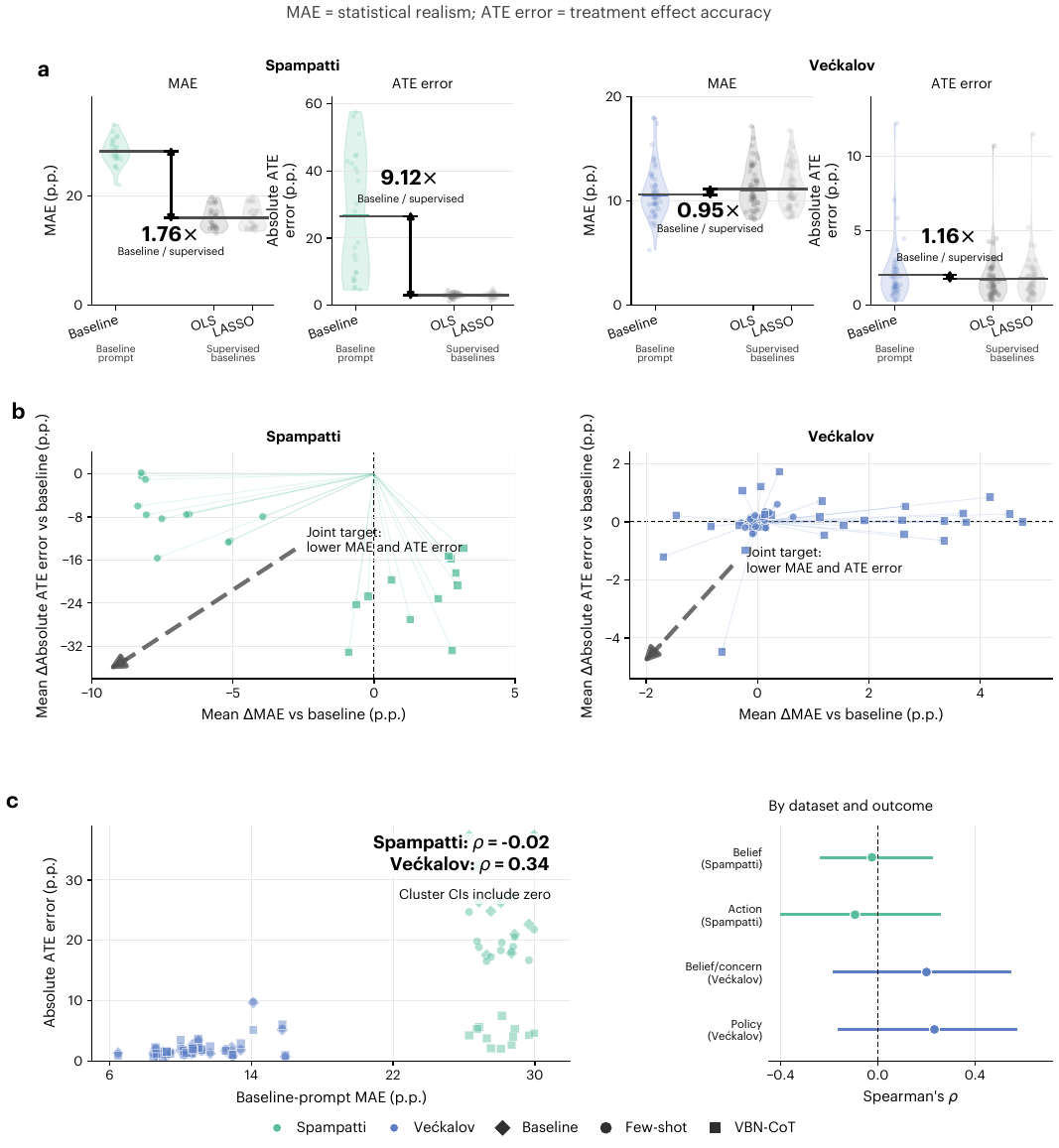}
  \caption{Replication analyses of the relationship between statistical realism and treatment-effect accuracy. Results are similar with Fig.~\ref{fig:fig2} but from two additional cross-national experiments: Spampatti et al. and Većkalov et al. \textbf{a}. Comparison of baseline LLM prompting and supervised baselines. The relative performance of LLMs and supervised models differed across datasets: LLMs showed larger errors in both MAE and ATE error in Spampatti et al., whereas their performance was comparable to supervised models in Većkalov et al. \textbf{b}. Changes in MAE and ATE error after prompt enhancement relative to baseline prompting. The horizontal and vertical axes indicate changes in MAE and ATE error, respectively. Points in the lower-left region indicate simultaneous reductions in both errors. \textbf{c}. Association between baseline statistical realism and treatment-effect accuracy across countries. Each point represents a country-level observation. Spearman correlations were estimated separately for each dataset, with confidence intervals accounting for country-level clustering. Across datasets, lower MAE was not consistently associated with lower ATE error. Note: We excluded perceived scientific consensus in the Većkalov dataset in this analysis because it was the construct directly targeted by the consensus-communication intervention, rather than a downstream outcome comparable to belief/concern and policy support.}
  \label{fig:s2}
\end{figure}

\begin{figure}
  \centering
  \includegraphics[width=0.9\linewidth]{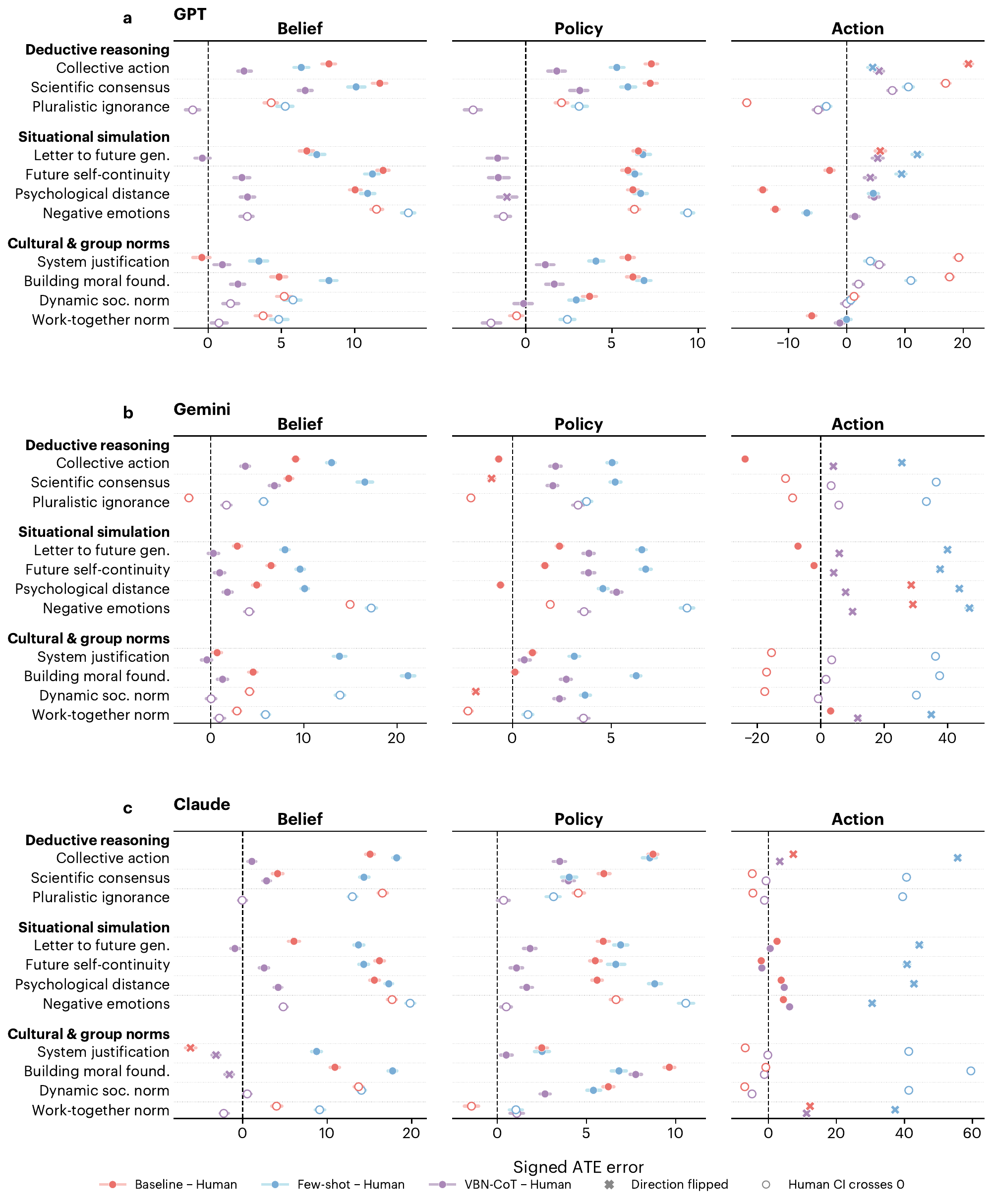}
  \caption{Intervention-level treatment-effect bias across interventions and prompting strategies. Each point represents the signed ATE error ($\Delta$ATE = predicted ATE - observed ATE) for one intervention under baseline (red), few-shot (blue), or VBN-CoT (purple) prompting. Positive values indicate overestimation and negative values indicate underestimation of treatment effects. \textbf{a}. GPT. \textbf{b}. Gemini. \textbf{c}. Claude. Crosses indicate interventions for which the estimated treatment-effect direction was opposite to the observed human effect, and open circles indicate interventions whose human ATE 95\% confidence intervals crossed zero. The three interventions selected for perturbation analyses in Fig.~\ref{fig:fig4} (scientific consensus, dynamic social norm, and future-self continuity) represent distinct theoretical categories and different ranges of treatment-effect bias. The results show substantial heterogeneity in treatment-effect errors across interventions, outcomes, and prompting strategies.}
  \label{fig:s3}
\end{figure}

\begin{figure}
  \centering
  \includegraphics[width=\linewidth]{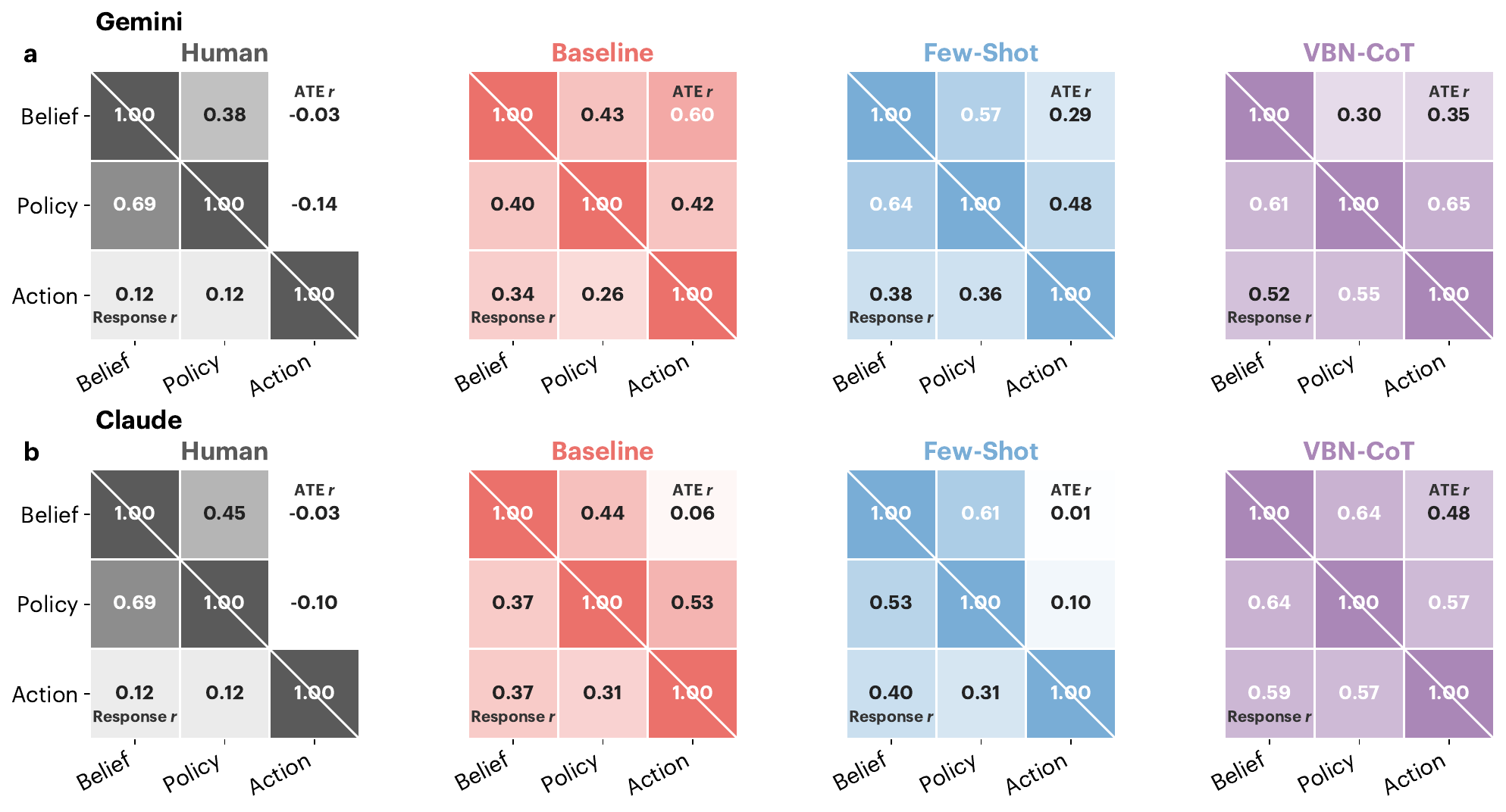}
  \caption{LLM simulations alter the association structure among behavioral outcomes.\textbf{a}. Gemini. b. Claude. Results for GPT are provided in Fig.~\ref{fig:fig4}b in the main text.  Lower triangles show individual-level response correlations. Upper triangles show intervention-level ATE correlations. Human data show weak associations between attitudinal and behavioral outcomes, whereas LLM-generated data exhibit stronger coupling, particularly under inference-based prompting. Like the results of GPT, these altered association patterns suggest that LLM simulations may not preserve the relationship structure between attitudes and behaviors observed in human data.}
  \label{fig:s4}
\end{figure}

\begin{figure}
  \centering
  \includegraphics[width=\linewidth]{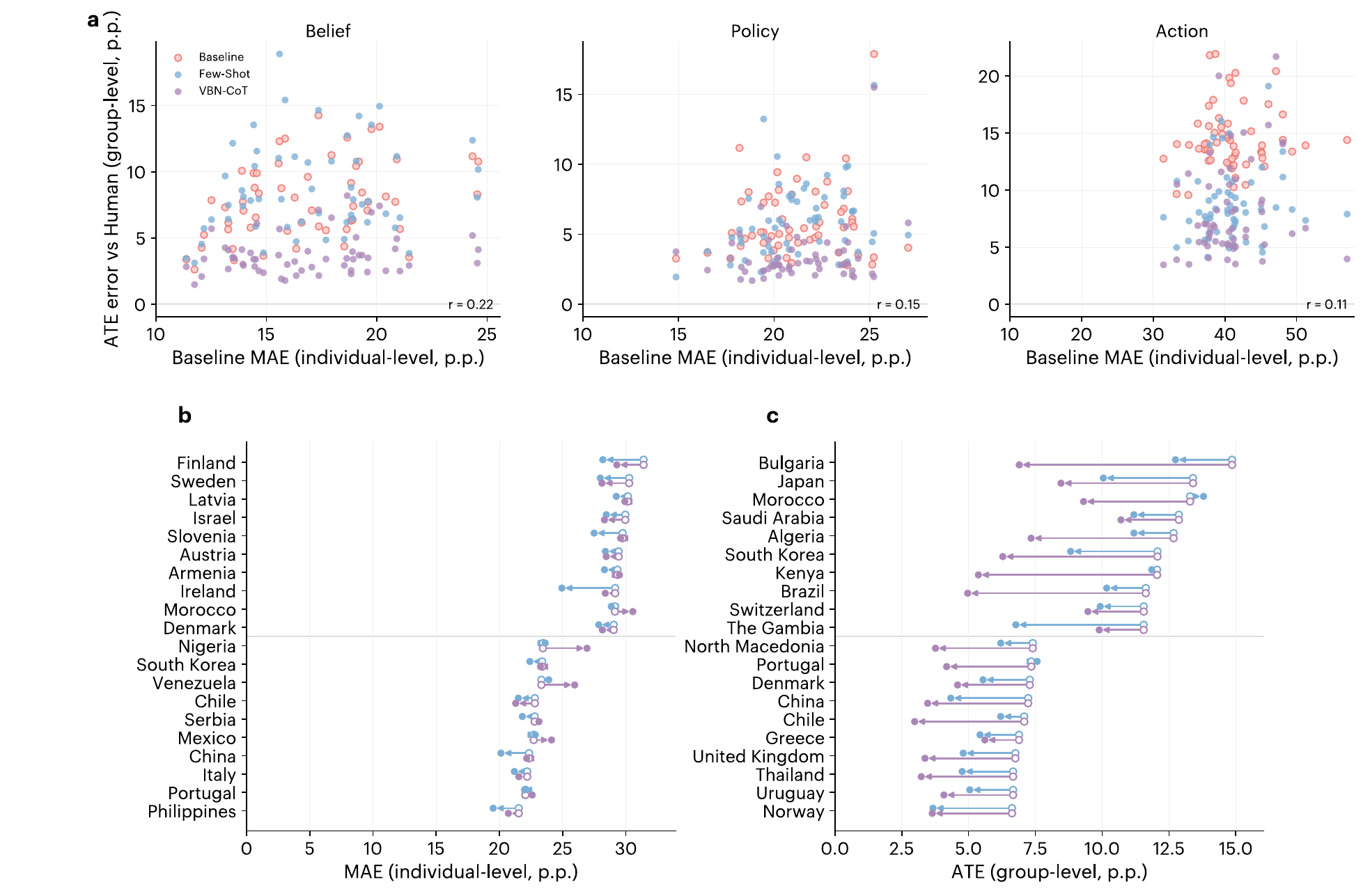}
  \caption{Divergence of country-level descriptive and causal error and their inequality profiles for \textbf{GPT}. \textbf{a}. Country-level scatterplots of baseline MAE against ATE error for each outcome. \textbf{b}. Countries ranked by MAE (top and bottom 10). \textbf{c}. Countries ranked by ATE error (top and bottom 10).}
  \label{fig:s5}
\end{figure}

\begin{figure}
  \centering
  \includegraphics[width=\linewidth]{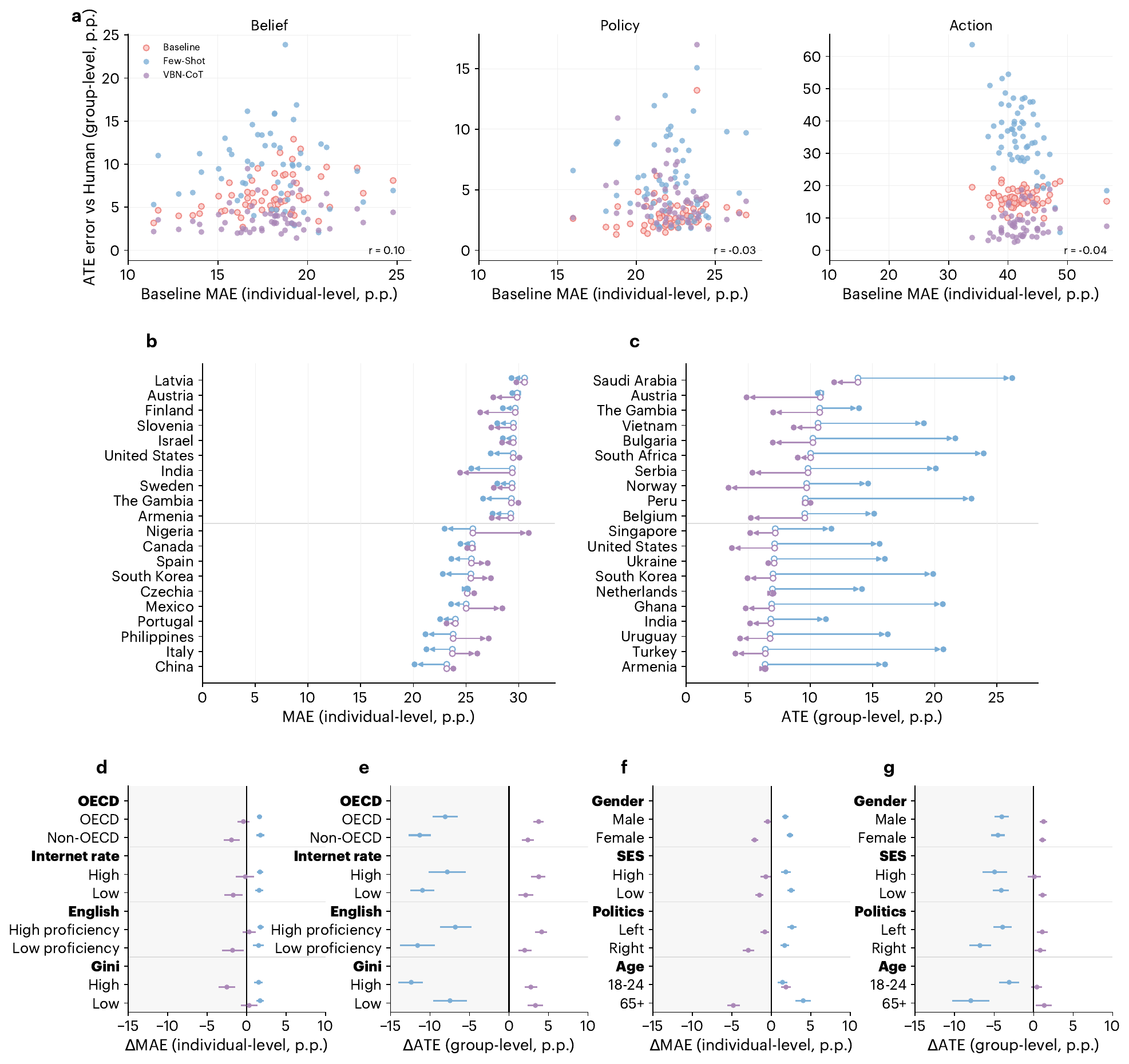}
  \caption{Divergence of country-level descriptive and causal error and their inequality profiles for \textbf{Gemini}. \textbf{a}. Country-level scatterplots of baseline MAE against ATE error for each outcome. \textbf{b}. Countries ranked by MAE (top and bottom 10). \textbf{c}. Countries ranked by ATE error (top and bottom 10). \textbf{d}. Between-group differences in MAE by country characteristics. \textbf{e}. Between-group differences in ATE error by country characteristics. \textbf{f}. Between-group differences in MAE by demographic characteristics. \textbf{g}. Between-group differences in ATE error by demographic characteristics.}
  \label{fig:s6}
\end{figure}

\begin{figure}
  \centering
  \includegraphics[width=\linewidth]{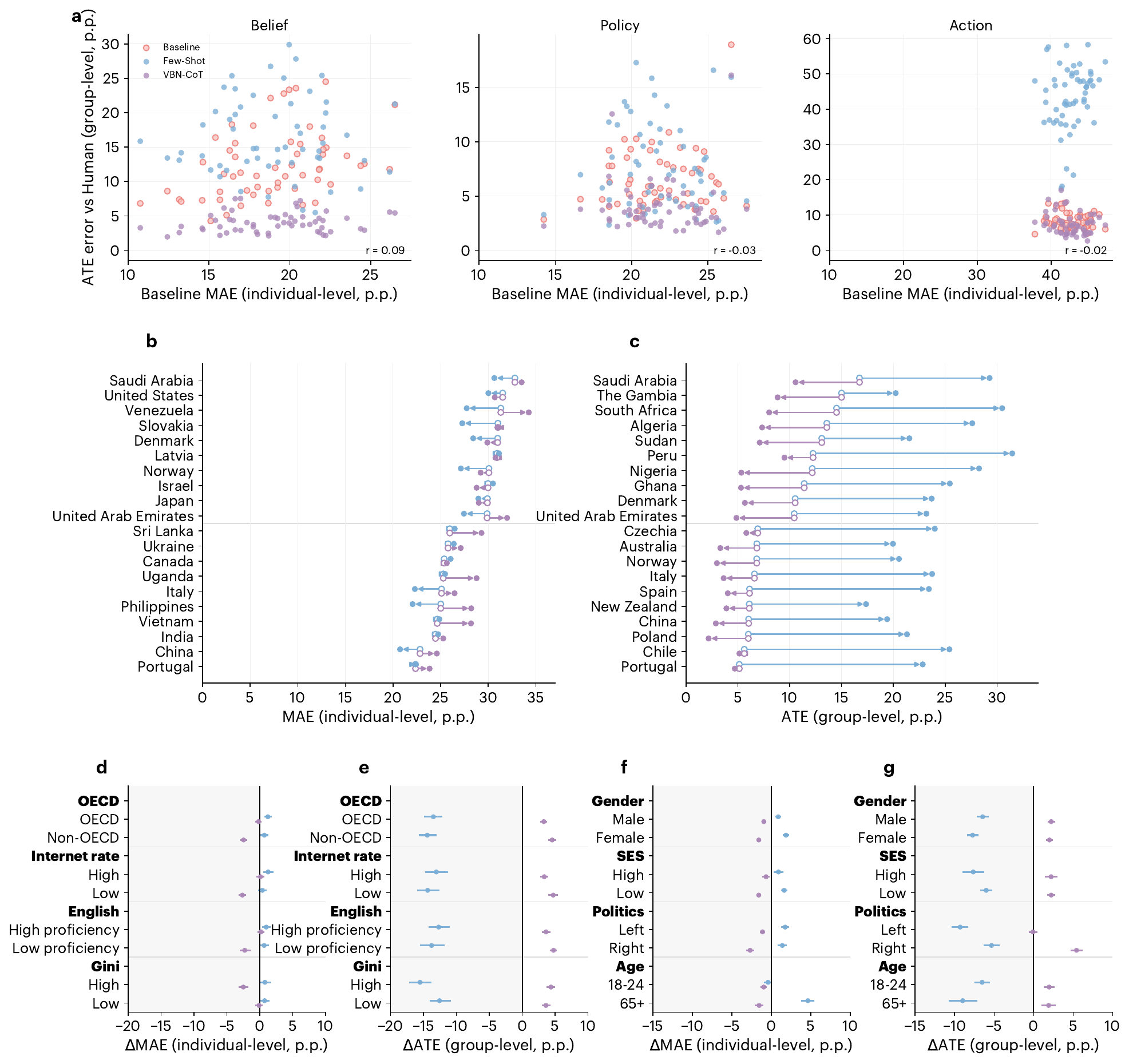}
  \caption{Divergence of country-level descriptive and causal error and their inequality profiles for \textbf{Claude}. \textbf{a}. Country-level scatterplots of baseline MAE against ATE error for each outcome. \textbf{b}. Countries ranked by MAE (top and bottom 10). \textbf{c}. Countries ranked by ATE error (top and bottom 10). \textbf{d}. Between-group differences in MAE by country characteristics. \textbf{e}. Between-group differences in ATE error by country characteristics. \textbf{f}. Between-group differences in MAE by demographic characteristics. \textbf{g}. Between-group differences in ATE error by demographic characteristics. }
  \label{fig:s7}
\end{figure}

\end{document}